\documentclass[journal]{IEEEtran}
\usepackage{amsmath,amsfonts}
\usepackage{algorithm}
\usepackage{array}
\usepackage[caption=false,font=normalsize,labelfont=sf,textfont=sf]{subfig}
\usepackage{textcomp}
\usepackage{stfloats}
\usepackage{url}
\usepackage{verbatim}
\usepackage{graphicx}
\usepackage{cite}
\usepackage{orcidlink}
\newcommand{\tinyskip}{\vspace{1pt}}

\newcommand{\mypar}[1]{\tinyskip\tinyskip\noindent\textbf{#1.}\xspace}
\newcommand{\sysname}{RELOAD\xspace}
\newcommand{\maml}{MAML\xspace}
\newcommand{\buffer}{PER\xspace}
\newcommand{\commdb}{SQL Server\xspace}
\newcommand{\robmax}{2.4\xspace}

\newcommand{\effmax}{3.1\xspace}

\usepackage{hyperref}
\usepackage{graphicx}
\usepackage{textcomp}
\usepackage{xcolor}
\usepackage{enumitem}
\usepackage{pifont}
\usepackage{tabularx}
\usepackage{algpseudocode}
\usepackage{booktabs}
\usepackage{float}
\usepackage{multirow}
\usepackage{xspace}
\usepackage{subfig}
\usepackage{comment}
\usepackage{tablefootnote}
\usepackage{makecell}
\usepackage[table]{xcolor}  
\begin{document}

\title{RELOAD: A Robust and Efficient Learned Query Optimizer for Database Systems}

% \author{IEEE Publication Technology,~\IEEEmembership{Staff,~IEEE,}}
\author{Seokwon Lee~\orcidlink{0009-0003-8251-0716}\IEEEmembership{, Graduate Student Member,~IEEE}, Jaeyoung Sim~\orcidlink{0009-0007-0308-5969}\IEEEmembership{, Graduate Student  Member,~IEEE}, Sihyun Kim~\orcidlink{0009-0005-3853-618X}\IEEEmembership{, Graduate Student  Member,~IEEE}, Yuhsing Li~\orcidlink{0009-0006-8760-6575}\IEEEmembership{, Graduate Student  Member,~IEEE}, Yiwen Zhu~\orcidlink{0009-0005-6857-7505}~\IEEEmembership{, Member,~IEEE} and Kwanghyun Park~\orcidlink{0000-0003-0757-2725}~\IEEEmembership{, Member,~IEEE}
\thanks{Corresponding Author: Kwanghyun Park.}
\thanks{Seokwon Lee, Jaeyoung Sim, Sihyun Kim, Yuhsing Li and Kwanghyun Park are with Yonsei University BDAI Lab, Seoul 03722, South Korea
(e-mail: \{guguri, jaeyoung.sim, sihyun.kim, yuhsing.li, kwanghyun.park\}@yonsei.ac.kr).
}
\thanks{Yiwen Zhu is with Microsoft Gray Systems Lab, Redmond, USA
(e-mail: yiwzh@microsoft.com)}
\thanks{This work is currently under review.}
}

\markboth{Journal of \LaTeX\ Class Files,~Vol.~14, No.~8, August~2021}%
{Shell \MakeLowercase{\textit{et al.}}: A Sample Article Using IEEEtran.cls for IEEE Journals}


\maketitle

\begin{abstract}
Recent advances in query optimization have shifted from traditional rule-based and cost-based techniques towards machine learning-driven approaches. Among these, reinforcement learning (RL) has attracted significant attention due to its ability to optimize long-term performance by learning policies over query planning. However, existing RL-based query optimizers often exhibit unstable performance at the level of individual queries, including severe performance regressions, and require prolonged training to reach the plan quality of expert, cost-based optimizers. These shortcomings make learned query optimizers difficult to deploy in practice and remain a major barrier to their adoption in production database systems.
To address these challenges, we present \sysname, a robust and efficient learned query optimizer for database systems. \sysname focuses on (i) robustness, by minimizing query-level performance regressions and ensuring consistent optimization behavior across executions, and (ii) efficiency, by accelerating convergence to expert-level plan quality. Through extensive experiments on standard benchmarks, including Join Order Benchmark, TPC-DS, and Star Schema Benchmark, \sysname demonstrates up to \robmax{\small$\times$} higher robustness and \effmax{\small$\times$} greater efficiency compared to state-of-the-art RL-based query optimization techniques.
\end{abstract}


\section{INTRODUCTION}
\IEEEPARstart{T}{he} query optimizer is an important component of database management systems (DBMSs). The tireless efforts of human experts have made traditional query optimizers perform well~\cite{LEON}. Traditional query optimizers are mostly cost-based, but even the most well-crafted cost-based optimizers still partially rely on rules from experts. Even the best cost-based optimization tools cannot completely eliminate this dependence on predefined cost models and heuristics, which limits their ability to handle complex data and query structures.
% which limits their adaptability to complex or previously unseen (i.e., ad-hoc) queries. 
In addition, traditional query optimizers have limitations when they lack key factors such as statistical information, data distribution, index availability, and query complexity~\cite{JOB}. To overcome the limitations of cost-based optimizers, a machine learning-based query optimizer is proposed~\cite{NEO}.
Among these, reinforcement learning (RL) has emerged as a particularly promising approach, because its iterative evaluation and improvement of decision policies resemble the dynamic programming strategies traditionally used in query optimization~\cite{qo,iter-dp,mb-rl}. This conceptual similarity has inspired recent research that integrates RL into query optimization, enabling data-driven learning beyond handcrafted cost models~\cite{opt-join-drl,hands-free-qo}. 
This integration enables query optimizers to leverage the predictive and adaptive capabilities of RL to generate more efficient query plans without expert intervention. With advances in deep neural networks, deep RL has been employed to effectively learn both the environment and state space of various logical and physical plans \cite{hands-free-qo}. 
\begin{figure}[!t]
    \centering
    \subfloat[Inconsistent robustness]{        
        \includegraphics[width=0.22\textwidth]{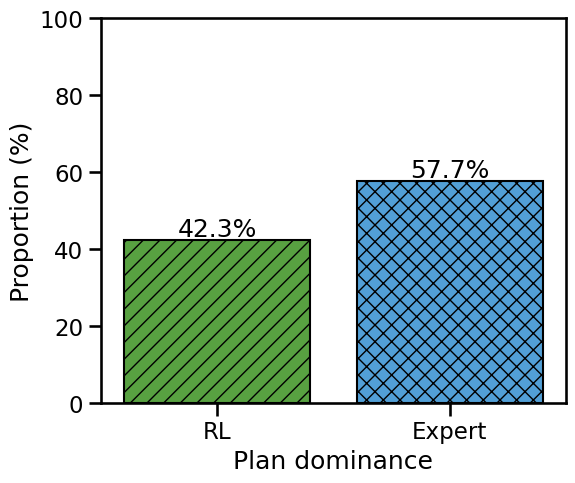}
        \label{fig:robustness}
    }
    \hfill
    \subfloat[Low efficiency]{
        \includegraphics[width=0.22\textwidth]{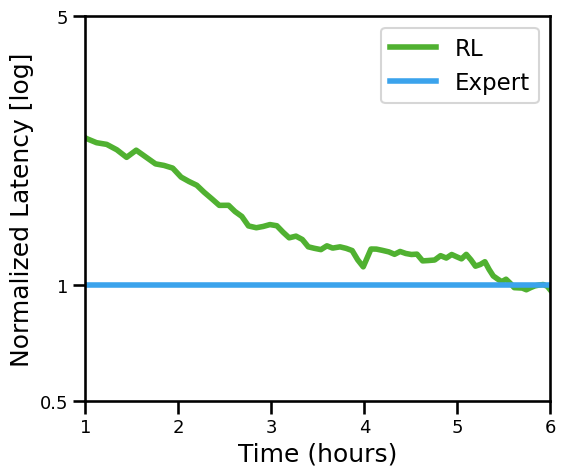}
        \label{fig:inefficiency}
    }
    \caption{Challenges encountered by an RL-based query optimizer on the Join Order Benchmark (JOB). Expert and RL are PostgreSQL and Balsa, respectively.
    (a) shows that RL is behind in about 58\% of the 26 test queries compared to the ones created by domain experts.
    (b) shows the time it takes for RL to catch up to the performance of the PostgreSQL optimizer.
    }
    % \kp{add expert in legend for (b)}}
    % \vspace{-6mm}
    \label{fig:challenges}
\end{figure}

% However, RL suffers from degraded performance when deployed in environments that deviate even modestly from those encountered during training ~\cite{challenges-real-world}. 
However, despite their potential, RL-based optimizers face significant challenges in replacing traditional optimizers due to inconsistent optimization stability and the excessive time required to reach a performance level comparable to expert-crafted plans in practical settings.
% This shortcoming stems from several core limitations of RL: its reliance on trial-and-error exploration makes it data-hungry and brittle; it is highly sensitive to distributional shifts in the environment; and its dependence on delayed, sparse, and often ambiguous reward signals significantly hinders effective credit assignment~\cite{rl-introduction, learning-TD}.
This shortcoming stems from several core limitations of RL: its reliance on trial-and-error exploration makes it sample inefficient and lacking robustness; it is highly sensitive to distributional shifts in the environment; and its dependence on delayed, sparse, and often ambiguous reward signals significantly hinders effective credit assignment~\cite{rl-introduction, learning-TD}.
% which attempts to maximize a reward in a given environment and is strongly dependent on the data collected.
While implementing state-of-the-art RL-based query optimizers such as Bao~\cite{Bao_QO}, Balsa~\cite{Balsa}, LOGER~\cite{LOGER}, and LIMAO~\cite{LIMAO} and benchmarking them against execution plans crafted by domain experts, we encountered several limitations (\autoref{fig:challenges}) in practical settings as follows. 

\begin{figure}[t!]
    \centering
     \subfloat[Plateau]{    
        \includegraphics[width=0.23\textwidth]{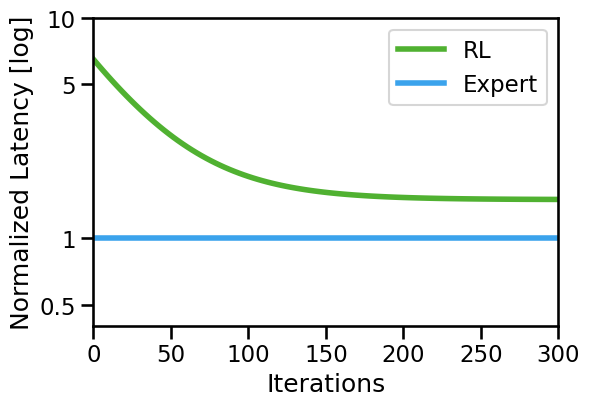}
        \label{fig:Plateau}
    }
    \hfill
    \subfloat[Rebound]{ 
        \includegraphics[width=0.23\textwidth]{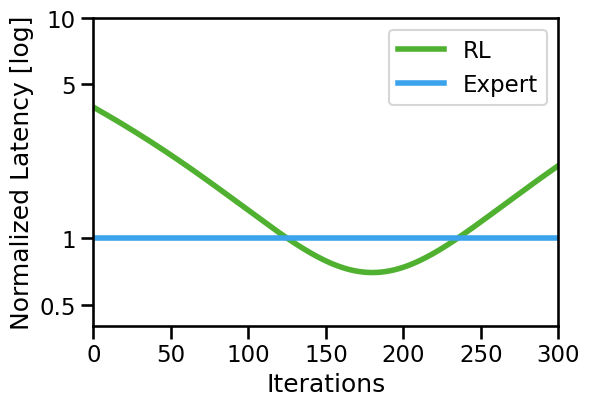}
        \label{fig:rebound}
    }
    % \caption{(a) and (b) illustrate two trends of performance regression observed in RL-based query optimizers, as tested on the JOB.}
    \caption{(a) and (b) illustrate two trends of performance regression observed in RL-based query optimizers, as tested on the JOB: convergence to suboptimal plans and performance regression, respectively.} \label{fig:PlateauNRebound}
    %\vspace{-4mm}
\end{figure}

\mypar{Robustness} 
Ensuring consistent robustness, where all queries reliably achieve optimal performance exceeding that of domain experts, remains a significant challenge.
As shown in \autoref{fig:robustness}, the RL-based optimizer fails to outperform expert plans in 57.7\% of the test queries, revealing its inconsistent robustness  across queries within the same workload. While most learned query optimizers report aggregate workload performance as their primary evaluation metric, this approach can obscure significant regressions on individual queries or corner cases. In contrast, this work targets \textit{query-level robustness}, aiming to ensure consistent and stable performance improvements across all individual queries. This objective imposes a substantially higher standard than aggregate metrics, as it requires optimizers to maintain stable performance effectively across diverse query patterns without sacrificing performance on any subset.
% \yz{can we emphasize here a little bit more saying that our requirement is even higher compared to the ones reporting aggregated performance where we want EVERY SINGLE query to have better performance?} 
Specifically, we want to avoid:
\begin{itemize} [left=0pt]
    \item \textbf{Convergence to suboptimal plans}: 
    % In some instances, the RL-generated plans converged prematurely, failing to reach the latency levels achieved by expert-crafted plans. This suggests that the model may have settled into local optima (see \autoref{fig:Plateau}).
    In some instances, the RL-generated plans converged prematurely, failing to reach the latency levels achieved by expert-crafted plans. The learning curve exhibits a \textit{plateau}-like shape (see \autoref{fig:Plateau}), where performance stagnates as the optimizer becomes trapped in local optima~\cite{rl-introduction}.
    \item \textbf{Performance regression}: 
    % We observe cases where, during training iterations, the performance of the RL-based optimizer deteriorated as iterations progressed. 
    % \SL{This may be related to difficulties in credit assignment, where beneficial early decisions are not consistently reinforced, which can result in regression or even a rebound effect (see \autoref{fig:rebound}).}
    We also observe cases where, during training iterations, performance initially improves but later deteriorates (see \autoref{fig:rebound}). This \textit{rebound} effect arises primarily from the credit assignment problem in sparse and delayed reward settings~\cite{learning-TD}, where beneficial early decisions are not consistently reinforced.
\end{itemize}

\mypar{Efficiency} 
Another challenge is the inefficiency in learning to reach a sufficient level of performance and in the design of the RL training process. The inefficiency appears primarily as an increase in the time required to run the total workload to reach the performance level of an expert. The underlying causes include not only the scarcity of training data but also the lack of data that effectively facilitates learning. In real-world systems, such inefficiencies can significantly impact overall execution time, which is a critical performance metric to ensure system efficiency and user satisfaction (see \autoref{fig:inefficiency}).
This inefficiency reflects a \textit{slow convergence} problem inherent in RL-based query optimizers, where convergence requires extensive interactions due to the numerous decisions involved in generating query plans, from logical to physical operators.

Overall, these challenges related to inconsistent robustness and slow convergence often result in noticeable gaps between simulated training environments and real-world deployments, which complicates the practical adoption of RL-based strategies.

% \mypar{Generalization versus customization}
% A general model, trained with data from diverse sources, offers the advantages of learning transfer between different workloads, adaptability to changes such as data size, and a potentially ``warmed-up" state for new workloads, enhancing early-stage performance~\cite{transfer-learning}. However, a one-size-fits-all approach often fails if specific workload characteristics are not adequately captured in the RL model's context or state. On the other hand, training a model for a specific query and fine-tuning it demands extensive data specific to that query. Although such models usually achieve better performance with longer training periods, they may lead to scalability and maintainability issues due to the need for numerous datasets~\cite{maintainability}.

% These challenges underscore the difficulty of simulating real-world systems with RL, given the vast and continuously evolving state and behavior spaces that are difficult to capture with limited samples for query optimization~\cite{challenges-real-world}. This complexity often leads to significant disparities between simulated training environments and real-world applications, further complicating the effective deployment of RL-based strategies. 

\mypar{Novelty \& Contributions}
In this work, we propose \sysname, a robust and efficient learned query optimizer designed to address the critical limitations of RL-based query optimizers---inconsistent robustness (i.e., \textit{Plateau}, \textit{Rebound}), and low efficiency (i.e., \textit{Slow convergence}). To the best of our knowledge, \sysname is the first to explicitly mitigate these query-level regressions. This contrasts with prior works~\cite{Balsa, LOGER, LIMAO, chen2024glo, weng2024eraser, li2025athena}, which primarily focused on aggregate performance.
To achieve this, rather than merely applying generic RL techniques, \sysname tailors two core mechanism specifically for the query optimization domain: knowledge retention to ensure robustness and knowledge transfer to accelerate efficiency.
\sysname operates as a complementary enhancement framework compatible with existing RL-based query optimizers, facilitating stable learning without altering their core architectures.
Specifically, our contributions are as follows:

\begin{itemize} [left=0pt]
\item \textbf{Implementation on a DBMS} (\autoref{sec:system-overview}): We integrate \sysname into both commercial and open-source DBMSs, demonstrating its applicability with existing RL-based frameworks, and release it as open-source\footnote{https://anonymous.4open.science/r/RELOAD}.

\item \textbf{Experience-aware knowledge retention} (\autoref{sec:retention}):
We introduce a specialized prioritized experience replay (PER) mechanism that extracts fine-grained learning signals instead of coarse-grained full plans. By prioritizing recent and hard-to-predict experiences, this module mitigates the sparse reward problem and prevents the optimizer from forgetting crucial high-cost signals.

\item \textbf{Complexity-aware knowledge transfer} (\autoref{sec:transfer}):
To accelerate convergence across diverse workloads, we devise a model-agnostic meta-learning (MAML) strategy integrated with novel workload partitioning policies. By grouping queries based on structural complexity (e.g., Halstead metrics) rather than naive arrival order, we enable an efficient initialization that accelerates convergence to surpass expert-level performance.
% we enable efficient initialization and rapid adaptation to ad-hoc queries.

\item \textbf{Query-level robustness formalization} (\autoref{sec:setup}):
We identify two distinct failure modes in RL-based query optimizers---convergence to suboptimal local minima (Plateau) and performance regression due to credit assignment failure (Rebound). We propose a metric to quantify these query-level performance regressions, overcoming the limitations of standard aggregate evaluations.

\item \textbf{Evaluation on diverse workloads} (\autoref{sec:evaluation}): Extensive experiments on benchmarks, including the Join Order Benchmark (JOB), TPC-DS, and Star Schema Benchmark (SSB), demonstrate \sysname's superiority, achieving 
\robmax{\small$\times$} higher robustness and \effmax{\small$\times$} greater efficiency compared to state-of-the-art approaches.
\end{itemize}
\section{BACKGROUND}
This section provides the learning principles behind \sysname, which addresses the key challenges of robustness and efficiency in query optimization. We first outline the concept of adaptive reinforcement learning, which enables an optimizer to iteratively refine its decision policies.
% learn effectively in evolving query environments where queries may exhibit unseen structures. 
We then describe two core modules that build upon this foundation: (1) knowledge retention, which enables robust learning by selectively reusing past experiences, and (2) knowledge transfer, which accelerates convergence through meta-learning by providing an informed initialization for various query structures.
% adaptation through meta-learning across related workloads.
These modules together form the basis of \sysname’s design for robust and efficient query optimization.

\subsection{Adaptive Reinforcement Learning}
Adaptive RL enables agents to operate effectively across complex optimization scenarios by retaining useful knowledge from past experiences and leveraging it to enhance learning efficiency. Its effectiveness relies on two complementary mechanisms: \textit{knowledge retention}~\cite{DRL_knowledge}, which ensures robustness by preventing the loss of previously acquired information, and \textit{knowledge transfer}~\cite{toward-crl}, which promotes efficiency by leveraging past knowledge to accelerate learning in new contexts. Together, these mechanisms enhance the adaptability of RL systems in practical DBMS environments.

\subsection{Knowledge Retention}
Knowledge retention is not just about storing experiences but about integrating accumulated knowledge. In reinforcement learning, this is particularly challenging because rewards are sparse and training is often unstable. Experience replay~\cite{self-improving} has been widely adopted to alleviate such issues and improve sample efficiency.

\textbf{Prioritized experience replay (PER)} \cite{per} has been used to achieve knowledge retention. In general, off-policy methods sample experiences evenly, whereas PER prioritizes by sampling high-importance experiences. PER follows a process of adding experiences, calculating priorities, sampling experiences, training the model, and updating priorities. This process reduces the loss of existing knowledge and ensures that important patterns are remembered effectively.
However, traditional PER assumes frequent rewards and stationary tasks, which makes it less effective for query optimization workloads that provide sparse and highly variable feedback. To address this, we design the replay mechanism for query optimization by weighting experiences according to their learning value, thereby improving robustness and stability.

\subsection{Knowledge Transfer}
Knowledge transfer~\cite{transfer-learning-rl} enables agents to leverage prior knowledge to enhance the learning efficiency of subsequent optimization tasks.
% allows agents to apply prior knowledge to new tasks for faster adaptation.
Meta-learning~\cite{learning-to-learn} is a commonly used method for knowledge transfer, and meta-reinforcement learning (meta-RL)~\cite{survey-meta-rl} is the application of the concept of meta-learning to RL.
Meta-RL facilitates the optimization process by providing robust initial parameters that serve as an informed starting point.
% Meta-RL helps agents adapt more efficiently to new tasks by learning robust initial parameters that generalize across a variety of workloads.

In query optimization, depending on the pattern or structure of the workload, each workload can be considered a different Markov decision process (MDP)~\cite{opt-join-drl, learning-state-rep}. 
Achieving rapid convergence toward expert-level performance requires initializing the model with parameters that effectively capture the underlying regularities across these MDPs. This improved initialization enhances both training efficiency and the quality of the resulting query plans.
% Rapid adaptation to a new workload thus requires initializing the model with parameters that generalize well across different MDPs. This improves both generalization and fine-tuning efficiency, ultimately leading to better performance.

\textbf{Model-agnostic meta-learning (MAML)}~\cite{maml} is one of the most common meta-learning algorithms for this purpose, and was specifically selected for \sysname because it offers superior initialization quality compared to standard transfer learning methods~\cite{generalization-maml}. 
While MAML facilitates efficient learning, it assumes predefined task boundaries, which do not naturally exist in query sets.
% While MAML enables fast adaptation, it assumes predefined task boundaries, which do not naturally exist in query workloads. Moreover, MAML does not prescribe how tasks should be formed.
We address this by applying a complexity-aware partitioning strategy to define meaningful task groups across, ensuring stable and accelerated convergence.
\section{SYSTEM OVERVIEW} \label{sec:system-overview}
\begin{figure*}[!t]
\centering
    \includegraphics[width=0.95\linewidth]{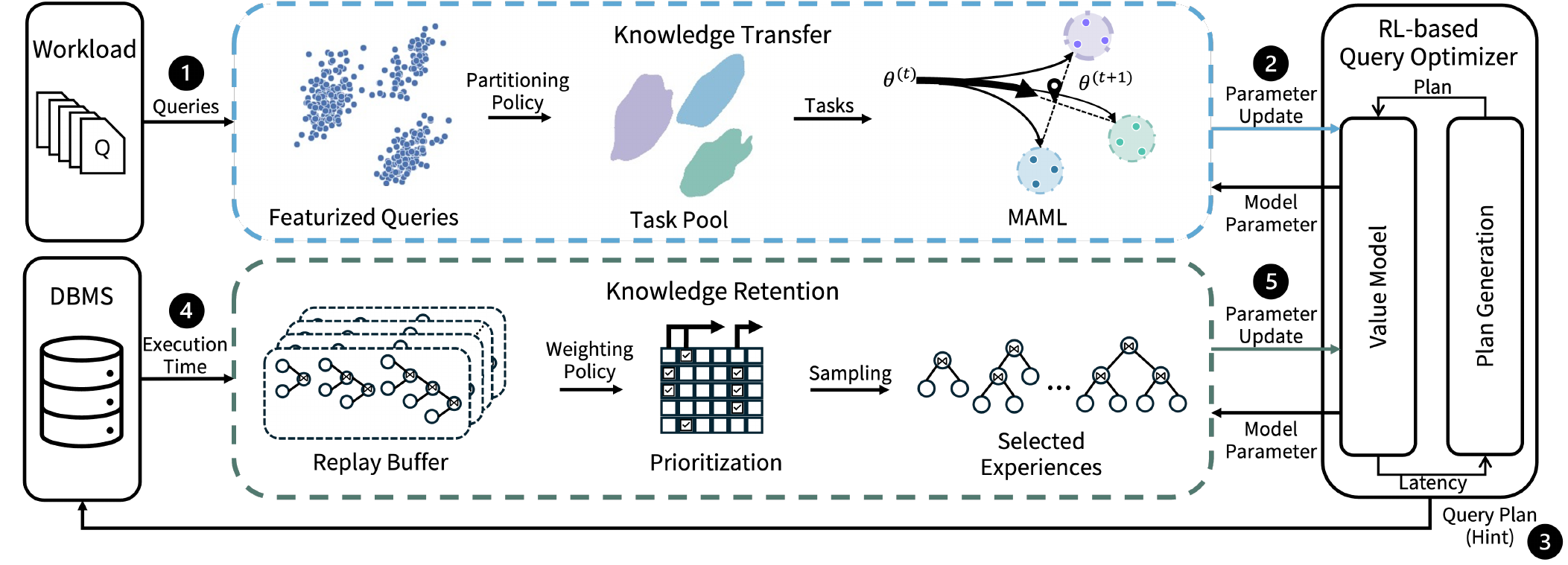}
    \caption{RELOAD integration into the RL-based Query Optimizer, with new components in dashed lines.}
    % \yz{try to make the fonts larger, same for Fig 4?}
    \label{fig:overall}
    % \vspace{-8pt}
\end{figure*}
\sysname is a robust and efficient RL-based query optimizer designed to ensure query-level performance stability and accelerated convergence toward expert-level optimization.
% quickly adapt to ad-hoc queries. 
As shown in \autoref{fig:overall}, the system integrates two core modules---Knowledge Transfer and Knowledge Retention---that complement each other to achieve both fast convergence and long-term robustness.

% \yz{This feels a bit weird that we start from number 3 instead of 1/2}

\mypar{Knowledge Transfer}
To reduce convergence time and achieve expert-level performance, \sysname first performs knowledge transfer through meta-learning. This process begins when training queries are received from a workload (\autoref{fig:overall} \ding{202}). Each query is featurized based on structural and execution-specific characteristics—such as the number of operators, estimated cost, and estimated cardinality—and then grouped into tasks using multiple partitioning policies. The most effective policy is selected using clustering metrics, ensuring that queries with similar complexity or behavior are trained together.

Once the task groups are defined, \sysname applies MAML to establish initial parameters across tasks. After convergence, the resulting meta-learned parameters are used to update the value model (\autoref{fig:overall} \ding{203}), providing an informed and transferable initialization that improves plan quality and accelerates convergence to outperform expert.

\mypar{Knowledge Retention} After initialization, \sysname proceeds with its training phase, where the DBMS generates and executes a query plan using the current value model initialized with meta-learned parameters (\autoref{fig:overall} \ding{204}). The execution latency returned from the DBMS serves as feedback to evaluate the quality of the generated plan (\autoref{fig:overall} \ding{205}). This feedback is then passed to the knowledge retention module, where the executed plan is processed through experience extraction (\autoref{fig:overall} \ding{206}).

Each experience, the basic learning unit in \sysname, corresponds to a featurized join-rooted subplan (i.e., a state extracted from an executed query plan). The extracted experiences are stored in a prioritized replay buffer and later resampled during training to reinforce valuable learning signals while mitigating performance regressions such as Plateau and Rebound.

% After initialization, \sysname proceeds with its training phase, where it continually adapts to incoming queries. For each query, a plan is generated and executed by the DBMS using the current value model initialized with meta-learned parameters (\autoref{fig:overall} \ding{204}). The execution latency returned from the DBMS serves as feedback, which is used to evaluate the quality of the generated plan (\autoref{fig:overall} \ding{205}).
% This feedback is then passed to the knowledge retention module. The executed plan is \SL{processed through experience extraction, where experiences derived from the plan are featurized and stored in a replay buffer.}
% During training, prioritized experiences are sampled from the replay buffer based on their computed weights. The sampled experiences are then used to update the value model (\autoref{fig:overall} \ding{206}), \SL{reinforcing valuable learning signals while mitigating performance regression phenomena such as Plateau and Rebound.}

% \yz{Can we add some discussion to highlight the methodological difference between LIMAO and this work?}
\section{ROBUSTNESS VIA KNOWLEDGE RETENTION} \label{sec:retention}
The knowledge retention module in the \sysname framework is designed to tackle two key performance challenges—\textbf{Plateau} and \textbf{Rebound}. 
An overview of the module’s role and significance within the framework is provided in \autoref{sec:retention:overview}, 
followed by detailed explanations of its two main components: experience extraction (\autoref{sec:retention_qd}) and PER (\autoref{sec:retention_per}). 
PER selectively reuses informative past experiences to recover from local optima (Plateau) and employs temporal-difference (TD) error~\cite{learning-TD} weighting to mitigate credit assignment issues (Rebound) by emphasizing experiences with high prediction errors, thereby reinforcing beneficial decisions under sparse reward conditions.
% stabilize learning under sparse rewards (Rebound), ensuring robust and consistent performance over time. Together, these components enable the optimizer to retain and prioritize impactful learning experiences, ensuring robust and consistent performance over time.

% The knowledge retention module in the \sysname framework is designed to tackle two key performance challenges: \textbf{Plateau} and \textbf{Rebound}. Plateau refers to cases where the optimization process prematurely converges to suboptimal policies, while Rebound occurs when a policy that initially performs well later deteriorates due to unstable credit assignment under sparse and delayed rewards. An overview of the knowledge retention module's role and significance within the framework is provided in \autoref{sec:retention:overview}. To address the challenges described above, the module integrates \SL{experience extraction} (\autoref{sec:retention_qd}) and \textit{prioritized experience replay (PER)} (\autoref{sec:retention_per}). Together, these components enable the optimizer to retain and prioritize impactful learning experiences, ensuring robust and consistent performance over time.

% \setlength{\abovecaptionskip}{-7pt}  % 그림과 캡션 사이 위쪽 여백
\begin{figure}[!tp]
\centering
    \includegraphics[width=0.95\columnwidth]{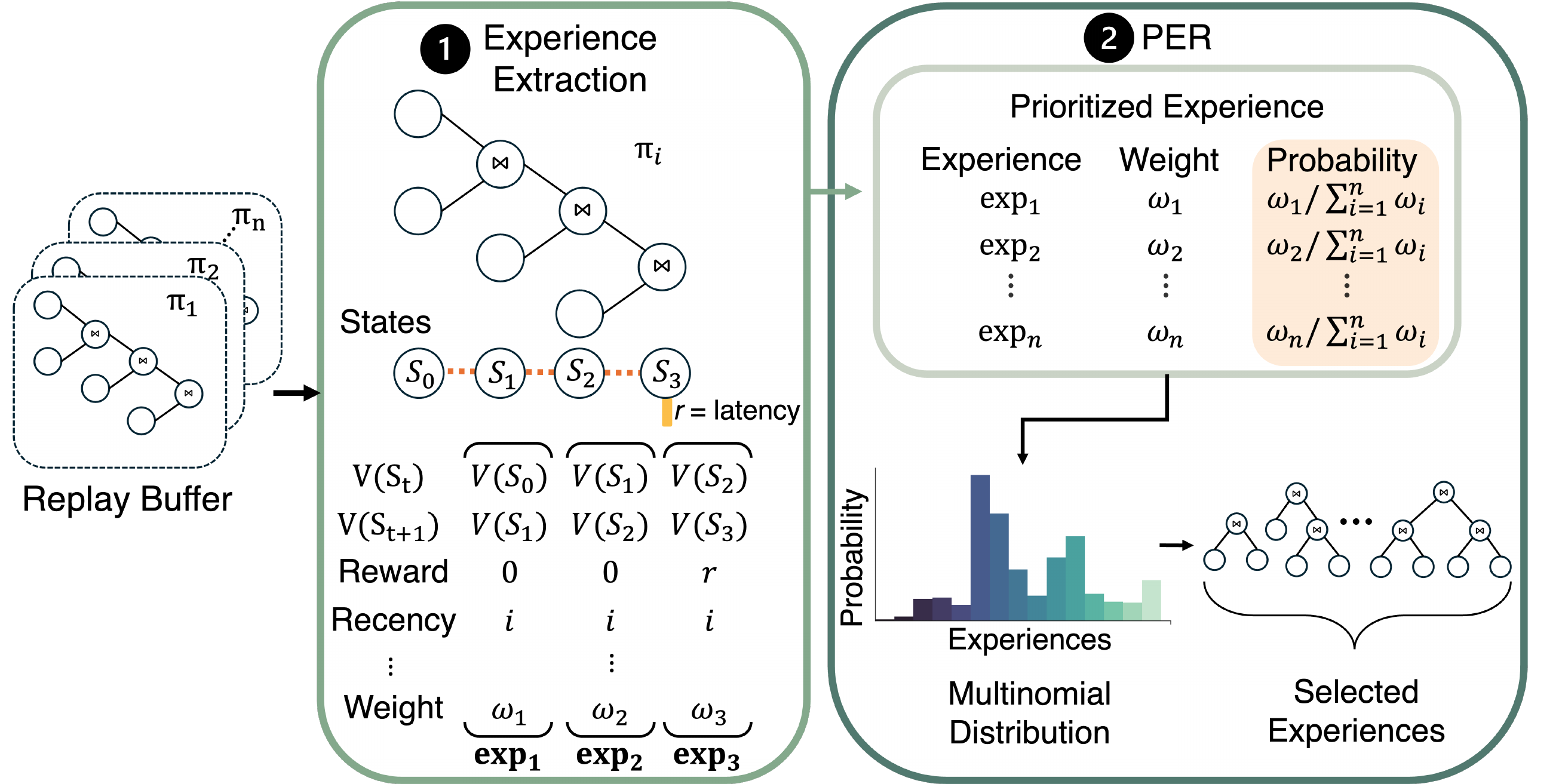}
    \caption{PER-based Knowledge Retention in \sysname. \ding{202} Experience Extraction gathers structural and runtime features from states. \ding{203} PER prioritizes the sampling of experiences from the replay buffer.} \label{fig:knowledge_retention}
    % \vspace{-5mm}
\end{figure}

\subsection{Overview of Knowledge Retention Process} \label{sec:retention:overview}
\autoref{fig:knowledge_retention} illustrates how the replay buffer is utilized for knowledge retention in \sysname. Starting from the left, the replay buffer maintains experiences derived from executed query plans. Each query plan is processed through \textbf{experience extraction}, where meaningful fragments of the plan are identified, featurized, and stored as experiences. This process (see \autoref{fig:knowledge_retention} \ding{202}) facilitates fine-grained learning by extracting and focusing on critical features, including structural attributes, execution metrics, and temporal information.

%\autoref{fig:knowledge_retenion} illustrates the process of leveraging the replay buffer for knowledge retention in the \sysname. Starting from the left, the replay buffer serves as a repository of past query execution plans and their sub-plans. Each query plan is decomposed into smaller, reusable sub-plans, which represent specific operations or subsets of the query. This \textbf{query decomposition} (\autoref{fig:knowledge_retenion}(a)) enables fine-grained learning by focusing on key features such as structural attributes, execution metrics, and temporal information.

In \autoref{fig:knowledge_retention} \ding{203}, the extracted features from experiences are then used in \textbf{PER}, where priority weights are calculated based on a defined weighting policy. The weighting policy ensures that the most impactful and relevant experiences are selected for training. These priority weights are normalized into probabilities, creating a multinomial distribution. Experiences are then stochastically sampled from this distribution, giving higher-priority experiences a greater likelihood of selection while still allowing for exploration of less frequently selected experiences. %ensuring that higher-priority sub-plans are more likely to be selected while still allowing for exploration of other sub-plans.

\subsection{Experience Extraction} \label{sec:retention_qd}

Experience extraction serves as the foundation for robustness by converting complex query execution plans into granular experiences. 
Each experience encapsulates structural and runtime features associated with subplans of the execution, enabling the optimizer to capture fine-grained variations across queries. 
% This process provides a more flexible basis for knowledge integration and improves adaptability across different query scenarios.
This process provides a more precise basis for knowledge retention and improves optimization reliability for diverse query structures.
While understanding the total cost at the plan level is essential, analyzing intermediate states within the plan provides critical insights that underpin the robust policy refinement aspect of \sysname. Each intermediate state in the MDP corresponds to a join-rooted subplan, capturing partial progress toward the final plan. By learning from these states, the optimizer can address fine-grained variations in execution characteristics and improve decision-making at a granular level. For practicality, we extract only the states that appear in the actual plan executed by the DBMS, focusing on subtrees rooted at join operators in the final plan. This design choice prevents the search space from becoming too large and allows the learning process to focus only on factors that are relevant to execution. To guide the prioritization within PER, we extract specific informative attributes from each experience that help assess its learning value beyond the basic state representation. In this step, the following information can be collected from each experience:
\begin{itemize} [left=0pt]
    \item \textit{Structural features}:  These features include the logical operators present in the subplan, such as joins, filters, and aggregations. They also include the types of joins used (e.g., hash join, nested loop join), which determine the execution strategy and cardinality estimates of intermediate states that indicate expected data sizes.
    \item \textit{Execution-specific features}: It includes estimated resource usage metrics, such as CPU, memory, and I/O, which indicate the computational requirements of the subplan. It also includes data size metrics, including row count, which corresponds to the cardinality, and data volume, which can be approximated by multiplying the row count by the average row width. These metrics give a better understanding of the data processing characteristics of the experience.
    \item \textit{Temporal features}: Recency of query execution helps assess how recently an experience was observed and its relevance to maintaining stable performance.
    \item \textit{Predicted execution latency:} The estimated time required to execute the given subplan under the current workload and system conditions.
\end{itemize}

The information gathered from these experiences is leveraged by PER to guide the prioritization and retention of experiences, identifying precisely the knowledge gaps of the current model. By integrating these features, PER effectively leverages the most relevant and impactful experiences to improve the adaptability and robustness.

\subsection{PER with Weighting Policies} \label{sec:retention_per}
To improve robustness and adaptability, RELOAD adopts \textbf{prioritized experience replay (PER)}~\cite{per}, which emphasizes informative or underrepresented experiences to mitigate credit assignment and local optima issues. We define a \textbf{weighting policy} $\omega$ that computes the importance of each experience based on \textbf{recency} and \textbf{temporal-difference (TD) error}, and normalize these weights for stochastic sampling in the replay buffer.
\begin{table}[h]
    \centering
    \caption{The weighting policy for prioritization.}
    \label{tab:weighting_policy}
    \begin{tabular}{c}
    \toprule
    \textbf{Weighting policy ($\omega$)} \\
    \midrule
        Recency-based\\
        TD error (low) \\
        TD error (high) \\
        Recency-based + TD error\\
    \bottomrule
    \end{tabular}
\end{table}

\mypar{Weighting policy} We introduce four weighting policies (\autoref{tab:weighting_policy}), which are computed based on two important weighting factors: \textit{Recency-based}  and \textit{TD error-based} weightings~\cite{per}. These policies balance the importance of recent experiences and the magnitude of prediction errors.

\begin{itemize}[left=0pt]
\item \textit{Recency-based}: Recent experiences are assumed to better reflect current execution conditions. The weight $\tau$ for a experience stored at time $\tau_e$ is computed as:
\begin{align}
\tau = 1 - \frac{\tau_{\text{current}} - \tau_e}{T},
\end{align}
where $\tau_{\text{current}}$ is the current time and T is a normalization constant representing the maximum possible time difference. This ensures $\tau \in [0, 1]$ and gives higher scores to more recent experiences.
% \yz{the t is used later as index. Might be good to distinguish. And sometimes we are using t eq. (2) and sometimes we are using i (4) and sometimes no index (3).}
\item \textit{TD error-based}: TD error quantifies the discrepancy between the predicted and observed execution latencies, highlighting experiences requiring further learning~\cite{learning-TD}. The TD error is defined as:
\begin{align}
\delta_t = r_{t+1} + \gamma V(s_{t+1}) - V(s_t), \label{eq:td-error}
\end{align}
Here, $r_{t+1}$ is the reward from the database engine, defined as the negative execution latency. $s_t$ and $s_{t+1}$ represent the current and next subplan states, where the latter reflects an additional join. The value function $V(\cdot)$ estimates latency using the value model, and $\gamma$ is the discount factor. TD errors are normalized via min-max scaling with exponent $\alpha$ as a tunable hyperparameter:
\begin{align}
\hat{\delta} = \frac{\delta^\alpha - \delta_{\text{min}}^\alpha}{\delta_{\text{max}}^\alpha - \delta_{\text{min}}^\alpha}.
\end{align}

Two strategies are used in our experiments: \textit{TD error (low)} and \textit{TD error (high)}, which assign higher weights to samples with smaller or larger TD errors, respectively. This choice determines whether sampling prioritizes stable or challenging experiences, influencing learning stability.

% \yz{what is the difference between low/high? Maybe try to make the notation consistent as well. t, delta\_t, delta\_alpha, omega(i), hat omega, ... some have input and some don't.}
% \JY{Intuitively, a high TD error indicates samples that the model fails to predict accurately, whereas a low TD-error corresponds to samples that are well predicted.}

\item \textit{Combined weighting policy}:
A flexible policy integrates both factors with a tunable hyperparameter $\beta$:
\begin{align}
\omega_i = \beta \cdot \hat{\delta}_i + (1 - \beta) \cdot \tau_i. \label{eq:weight}
\end{align}
When $\beta=0$, only recency is used; when $\beta=1$, only TD error is used. Intermediate values allow for a weighted combination.

\end{itemize}

\begin{algorithm}[!t]
    \small
    \caption{Selective Experience Replay} \label{algo:replay_buffer}
    \begin{flushleft}
    \textbf{Input:} Replay buffer of executed plans $\mathcal{D}=\{\pi_1,\dots,\pi_M\}$, weighting policy $\omega$, replay budget $k$ \\
    \textbf{Output:} \emph{Selected Experiences $S$}
    \end{flushleft}
    \begin{algorithmic}[1]
    \State $S \gets \emptyset$
    \State $X \gets \emptyset$
       \ForAll{$\pi_i \in \mathcal{D}$}
             \State $\mathcal{X}_i \gets \textsc{ExtractExperiences}(\pi_i)$
            \ForAll{$\mathrm{exp}_{i,j} \in \mathcal{X}_i$}
                \State $s_{i,j} \gets \omega(\mathrm{exp}_{i,j})$
                \State $X \gets X \cup \{(\mathrm{exp}_{i,j},\, s_{i,j})\}$
            \EndFor
        \EndFor

    \State $Z \gets \sum s_{i,j}$ \Comment{Normalization constant}
\State $p(\mathrm{exp}_{i,j}) \gets s_{i,j} / Z$
\For{$n = 1$ \textbf{to} $k$}
    \State $\mathrm{exp}_n \sim p(\mathrm{exp})$ \Comment{Sample from $X$ according to $p$}
    \State $S \gets S \cup \{\mathrm{exp}_n\}$
\EndFor
    
    \State \textbf{Return $S$}
    \end{algorithmic}
\end{algorithm}

\mypar{Sampling strategy} Based on the weights defined in \autoref{eq:weight}, the priorities are normalized into a probability distribution:

\begin{align}
\hat{\omega}_i = \frac{\omega_i}{\sum_{j=1}^{N} \omega_j} \hspace{1cm} i=1,\dots,N.
\end{align}
Here, $N$ denotes the total number of experiences. Experiences are then sampled from a multinomial distribution with a fixed budget $k$. The selected experiences, denoted by $X$, are drawn as:
\begin{align}
X = (X_1, X_2, \dots, X_k) \sim \text{Multinomial}(k, \hat{\omega}_1, \dots, \hat{\omega}_N).
\end{align}
This strategy increases the likelihood of selecting high-importance experiences, thereby reinforcing the learning process with more impactful samples. In sum, \autoref{algo:replay_buffer} illustrates the process of generating the replay buffer using the weighting policy. Each experience, derived from a subplan, is assigned a priority using the selected weighting policy. Experiences are then sampled based on their normalized priorities until the budget is exhausted (lines 12-15).

\mypar{Implementation note}
Unlike standard PER~\cite{per}, \sysname stores only recent experiences without priority updates to reduce overhead, while recency-based weighting still emphasizes new samples.

In summary, this replay mechanism allows \sysname to efficiently focus on subplans that matter most for learning, enhancing both convergence and robustness. The replay of structural features of the subplan by PER also has the effect of enhancing the efficacy improvement of MAML in terms of feature reuse~\cite{toward-understanding-maml}.
\section{EFFICIENCY VIA KNOWLEDGE TRANSFER} \label{sec:transfer}

The knowledge transfer module in the \sysname framework is designed to accelerate convergence while reducing ramp-up time by providing an informed initialization. This module leverages \textbf{model-agnostic meta-learning (MAML)}, a flexible and powerful meta-learning technique that facilitates efficient learning by establishing high-quality initial parameters that can be fine-tuned for specific tasks~\cite{generalization-maml}. This section first introduces the overall knowledge transfer module of \sysname (\autoref{sec:transfer:overview}). Next, we discuss the task partitioning policies which are evaluated using a clustering metric to ensure optimal grouping (\autoref{sec:transfer:policies}). Finally, we detail the MAML-based meta-learning process, demonstrating how task-specific and cross-task parameters are refined to maximize convergence efficiency (\autoref{sec:transfer:loop}).

\begin{figure}[!tp]
    \centering 
    \includegraphics[width=0.95\columnwidth]{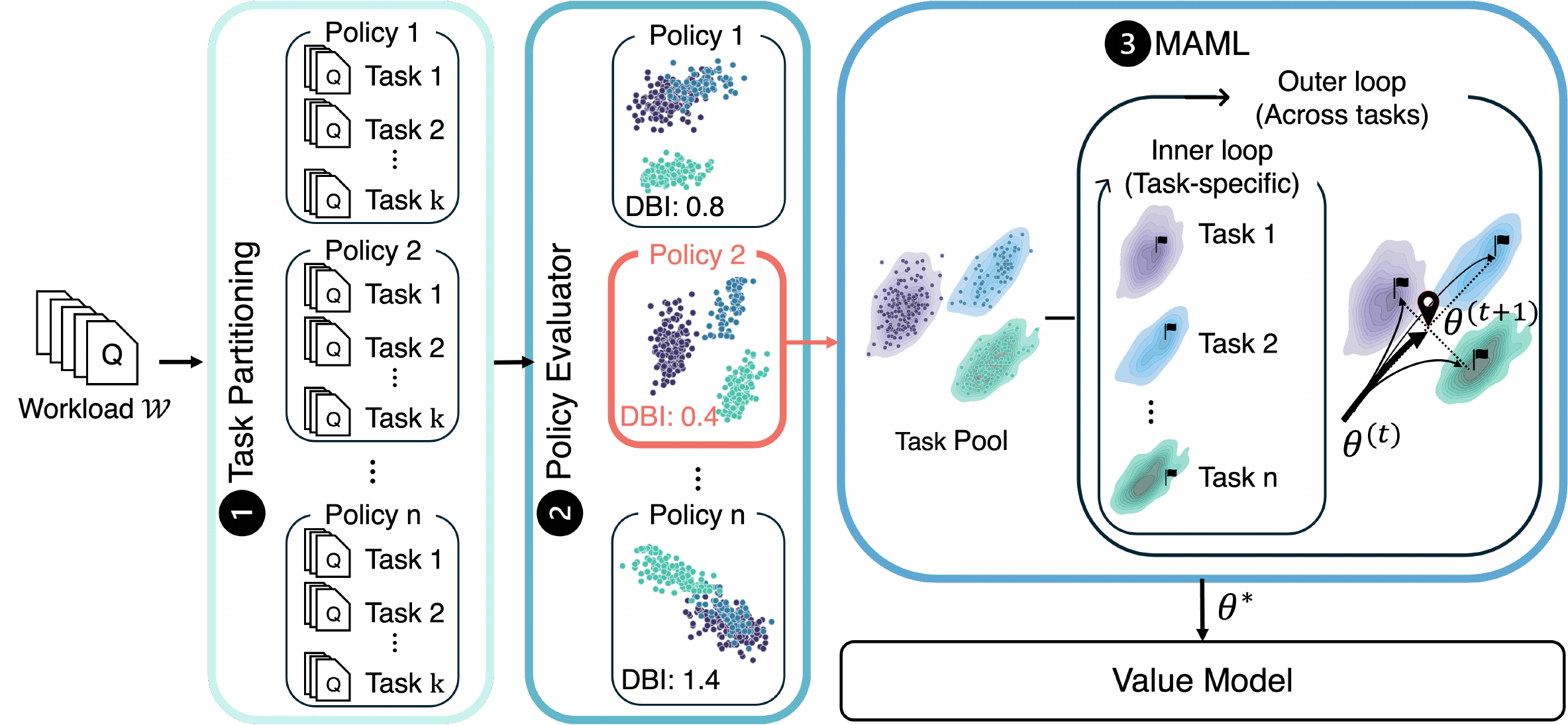} 
    \caption{MAML-based Knowledge Transfer in \sysname. \ding{202} Task Partitioning groups queries by policy. \ding{203} Policy Evaluator selects the best policy. \ding{204} Inner loop learns task-specific models; outer loop refines shared initialization across tasks.}\label{fig:knowledge_transfer}
    % \vspace{-4mm}
\end{figure}

\subsection{Overview of Knowledge Transfer Process} \label{sec:transfer:overview}
\autoref{fig:knowledge_transfer} illustrates the MAML-based process utilized in \sysname for knowledge transfer. The process starts with a workload $W$, partitioned into query groups (tasks) based on structural or execution features such as operator count or estimated cost. This \textbf{task partitioning} (\autoref{fig:knowledge_transfer} \ding{202}) enhances inner-loop refinement and outer-loop aggregation.

Queries can be grouped based on various criteria. For instance, Policy 1 may use structural query characteristics, and Policy 2 may focus on execution-specific metrics.
Multiple partitioning policies are evaluated by a \textbf{policy evaluator}, which selects the one achieving the best grouping quality (\autoref{fig:knowledge_transfer} \ding{203}).
Once the optimal policy (e.g., Policy 2) is selected, the corresponding groups of queries (tasks) are passed to the \textbf{MAML} process. During this process, the outer loop iterates repeatedly, with the inner loop being executed multiple times in each duration.

Finally, the task-specific parameters are refined through gradient updates in the inner loop, tailored to each task, and then aggregated in the value model during the outer loop (\autoref{fig:knowledge_transfer} \ding{204}). This nested process ensures that the model accelerates convergence across various tasks by improving the initial parameter $\theta$ through iterative optimization at both levels.

Finally, the task-specific parameters are refined through gradient updates in the inner loop, tailored to each task, and then aggregated in the value model during the outer loop (\autoref{fig:knowledge_transfer} \ding{204}). This nested process ensures that the model generalizes efficiently to new tasks by improving the initial parameter $\theta$ through iterative optimization at both levels.

\subsection{Task Partitioning Policy Selection} \label{sec:transfer:policies}
% \SL{Section 5.2 is kind of repetitive. Also Query classification is kind of misleading.}
To effectively apply MAML, workloads must be divided into tasks. Task partitioning relies on structural and execution-specific features of queries, such as:
\begin{itemize} [left=0pt]
    \item \textit{Structural features}: Logical operators, physical operators, and query complexity.
    \item \textit{Execution-specific features}: Estimated costs (CPU, memory, I/O), data size, and cardinality estimates.
\end{itemize}

Queries within a task should share similar features to ensure consistency in the inner-loop updates. At the same time, tasks should exhibit diversity to enable effective knowledge aggregation during the outer loop. To evaluate the quality of task partitioning produced by each policy, a clustering metric such as the \textbf{Davies-Bouldin Index (DBI)}~\cite{cluster-separation-measure} can be used~\cite{study-cluster-validity}. The DBI quantifies intra-task similarity and inter-task separation. The policy evaluator selects the policy with the lowest DBI, as this indicates the most suitable task partitioning. This ensures that:
\begin{itemize} [left=0pt]
    \item Queries within each task are highly similar, allowing consistent and effective inner loop updates.
    \item Tasks are sufficiently distinct, ensuring well-informed initialization across diverse workloads in the outer loop.
\end{itemize}

\begin{table}[h]
    \centering
    \caption{The partitioning policy for workload.}
    \label{tab:partitioning_policy}
    \begin{tabular}{c c}
        \toprule
        \textbf{Partitioning policy ($\mathcal{P}$)} & Features \\
        \midrule
        \multirow{3}{*}{Query complexity} & Halstead complexity measures \\ & Total number of operators \\ & Estimated query cost \\
        \midrule
        Data complexity & Estimated rows \\    
        \bottomrule
    \end{tabular}
    %\vspace{-4mm}
\end{table}

\mypar{Partitioning policy} We introduce four partitioning policies (\autoref{tab:partitioning_policy}) designed to classify queries based on structural and execution-specific features:
\begin{itemize}[left=0pt]
    \item \textit{Halstead complexity measures}~\cite{measuring-query-complexity}:  
    Halstead complexity measures are a quantitative metric for evaluating query complexity, adapted from metrics originally designed for programming languages. Each component of the equation corresponds to elements within SQL: 
    
    {\setlength{\abovedisplayskip}{0pt}
    \begin{align*}
        N &= \text{Total number of operands.} \\
        \eta_1 &= \text{Number of distinct operators.} \\
        \eta_2 &= \text{Number of distinct operands.}
    \end{align*}}
    The query complexity is then computed as:
    \[ 
    \text{Query complexity} = \frac{\eta_1}{2} \times \frac{N}{\eta_2} \times \log_2(\eta_1 + \eta_2).
    \]
    For example, in the query:
    {
    \begin{verbatim}
    SELECT MIN(t.title) FROM title AS t;
    \end{verbatim}
    }

    \texttt{SELECT}, \texttt{MIN}, and \texttt{FROM} are operators, while \texttt{t} and \texttt{title} are operands (\(N\)).
    
    \item \textit{Number of operators}:  
    The number of operators is a simple yet effective indicator of query complexity. This metric is derived directly from the Halstead complexity measures, focusing on the count of logical operations within the query.

    \item \textit{Estimated query cost}:  
    Query cost is calculated by database systems (e.g., PostgreSQL) based on resource usage, including CPU, disk I/O, and memory. Using the \texttt{EXPLAIN} command, we extract the \textsf{Total Cost} to represent the estimated query cost.

    \item \textit{Estimated rows}:  
    Cardinality, or the number of rows returned by operations, is a key factor in query optimization ~\cite{Survey_on_Advancing_DBMS_Query_Optimizer, JOB}. The estimated number of rows can be extracted from the \textsf{Plan Rows} value in the query plan generated by the \texttt{EXPLAIN} command.

    % \item \textit{Comprehensive Complexity}:  
    % This metric combines several features, such as the Halstead complexity, estimated query cost, and cardinality, to provide a holistic measure of query complexity.
\end{itemize}

% \yz{the algo 2 is not discussed in text}
\begin{algorithm}[t]
%\ContinuedFloat
\small
\caption{Partitioning Workloads with Optimal Policy}\label{algo:partitioning}
\begin{flushleft}
\textbf{Input:} Workload $\mathcal{W}$, Partitioning policy $\mathcal{P}$, Task set size $k$\\
\textbf{Output:} \emph{Task set}
\end{flushleft}
\begin{algorithmic}[1]
\State $\text{DBI}_{min} \gets +\infty$
\State $T_{best} \gets \emptyset$
\ForAll{\( p \in \mathcal{P}\)}
\State Tasks $T \gets \emptyset$
    \ForAll{\( q \in \mathcal{W} \)}
        \State $\text{score} \gets p(q)$ 
        
        \State $T \gets T \cup \{(q, \text{score})\}$
    \EndFor
    \State Sort $T$ by $\text{score}$ in ascending order
    \State $\hat{T} \gets \emptyset$
    \State $\text{size} \gets \lfloor \frac{| \mathcal{W} | }{k} \rfloor$
    \For{$ 1 \leq i < k$}
        \State $\hat{T} \gets \hat{T} \cup T[(i-1) \times \text{size} : i \times \text{size}]$
    \EndFor
    \State $\text{DBI} \gets \textsc{DaviesBouldinScore}(\hat{T})$
    \If {$\text{DBI}_{min} > \text{DBI}$}
        \State $\text{DBI}_{min} \gets \text{DBI}$
        \State $T_{best} \gets \hat{T}$
    \EndIf
\EndFor
\State \textbf{Return $T_{best}$}
\end{algorithmic}
\end{algorithm}

\mypar{Evaluating task partition policy} 
To classify tasks effectively, tasks must exhibit intra-task similarity and inter-task separation. To evaluate this, we use the DBI~\cite{cluster-separation-measure}:
\begin{equation*}
DBI = \frac{1}{k} \sum_{i=1}^{k} R_i,
\end{equation*}
where $R_i$ is the worst-case similarity between task $i$ and all other tasks, computed as:
\begin{gather}
    % \text{Minimize}\; DBI = \frac{1}{k} \sum_{i=1}^{k} R_i^p \quad\quad\forall p \in \mathcal{P} ,\\
    R_i =\max_{\substack{j = \{1,\dots, n\}, \\ i \neq j}} R_{ij} \quad \text{and} \quad  R_{ij} = \frac{\sigma_i + \sigma_j}{d_{ij}} ,\\
    \sigma_i = \frac{1}{|T_i|} \sum_{x \in T_i} d(x, c_i) \qquad T_i \subset
 \mathcal{W} ,\\
    d_{ij} = \max(\epsilon, d(c_i, c_j)) \qquad\qquad \epsilon > 0 ,\\
    \sum_{i=1}^{k} \vert T_i \vert = \vert \mathcal{W} \vert.\hspace{3cm}
\end{gather}

Each task is a cluster of query embeddings. $\sigma_i$ is the average distance between all query embeddings in the cluster. We use the center of cluster $c_i$ to find the distance between points in the cluster and the center, or the distance between clusters. We denote the distance between $c_i$ and $c_j$ as $d(c_i, c_j)$. Given a workload $\mathcal{W}$, we denote by $T_i$ each task created by partitioning the workload according to the policy. Different tasks must be a disjoint set, so the total number of queries in the tasks must equal the number of queries in the workload. A smaller DBI implies better clustering, and the policy that minimizes DBI is selected.
% \[
%     \min_{p \in \mathcal{P}} DBI_p
% \]

As described in \autoref{algo:partitioning}, the policy minimizing the DBI is selected to form meaningful tasks for MAML, ensuring high intra-task similarity and inter-task diversity. For practicality, we adopt a fixed number of equally sized tasks to ensure stable meta-training and balanced gradients~\cite{MetaLearning}, leaving adaptive grouping as future work.

% As described in \autoref{algo:partitioning}, each partitioning policy $p \in \mathcal{P}$ is evaluated by dividing the workload $\mathcal{W}$ into $k$ tasks and calculating the DBI for the resulting partition. The policy with the lowest DBI is chosen to ensure distinct clusters. Partitioning policies and their evaluation through DBI allow \sysname to create meaningful tasks for MAML, ensuring effective learning and generalization across diverse workloads. We perform this partitioning once before training begins. By integrating structural query features and embedding-based techniques, \sysname leverages both traditional and advanced methods to optimize task partitioning. While we extract query embeddings, we do not use them directly for clustering. Instead, they are used to evaluate partitioning policies via metrics like DBI. This approach improves interpretability and control, as embedding-based clusters can be hard to analyze. Using structured features such as query complexity or cost allows more transparent and adjustable task partitioning. 

\subsection{Meta-Learning and MAML in Query Optimization} \label{sec:transfer:loop}
The MAML process involves two key phases: the \textbf{inner loop}, which performs task-specific refinement, and the \textbf{outer loop}, which aggregates knowledge to establish a robust initialization across all tasks.

\mypar{Inner loop: Task-specific refinement}
The inner loop adapts the model to a specific task $T_i$ by performing gradient updates on the task-specific loss function $\mathcal{L}_{T_i}$. At iteration $t$, we refine the initial parameters $\theta^{(t)}$ into task-specific parameters $\theta_i^\prime$:
\begin{equation}
    \theta_i^\prime = \theta^{(t)} - \alpha \nabla_\theta \mathcal{L}_{T_i} (f_\theta).
\end{equation}

Here, $\alpha$ is the inner-loop learning rate, and $f_\theta$ is the model with initial parameters $\theta$. Task-specific refinement allows the model to effectively capture the distinct workload characteristics of task $T_i$.

\mypar{Outer loop: Cross-task aggregation}
The outer loop aggregates results from multiple tasks to learn a set of high-quality initial parameters $\theta^{(t+1)}$, by minimizing accumulated losses computed with task-specific parameters $\theta_i^\prime$:
\begin{equation}
\theta^{(t+1)} = \theta^{(t)} - \beta \nabla_\theta \sum_{i=1}^{N} \mathcal{L}_{T_i} (f_{\theta_i^\prime}).
\end{equation}
Here, $\beta$ is the outer-loop learning rate and $N$ is the total number of tasks. By training across diverse tasks, the outer loop ensures that the updated parameters $\theta^{(t+1)}$ serve as an informed initialization. As shown in \autoref{fig:knowledge_transfer}, it refines $\theta^{(t)}$ toward the optimal parameters that provide a reliable starting point for subsequent learning phases.
\section{Evaluation} \label{sec:evaluation}
In this section, we evaluate the performance of \sysname on different workloads using two database systems---PostgreSQL and a commercial DBMS (\commdb). We describe our experimental setup in \autoref{sec:setup}. First, we summarize the overall performance of \sysname using Workload Relative Latency (WRL)~\cite{LOGER, FOSS}, which measures the performance of a learned query optimizer relative to an expert optimizer across the entire workload. We then compare our approach with state-of-the-art methods---Bao, LOGER, Balsa and LIMAO---from two perspectives: \textit{Robustness} and \textit{Efficiency}. These metrics are separately evaluated for PostgreSQL and \commdb in \autoref{sec:macro}.

Finally, we conduct micro-experiments to show the effectiveness of the two components of \sysname---\textit{knowledge retention} and \textit{knowledge transfer}---with results reported in \autoref{sec:micro}. Our key findings on PostgreSQL are as follows:
\begin{itemize} [left=0pt]
    \item \sysname demonstrates consistent robustness across various workloads. Among all configurations, Balsa (vanilla) + \sysname achieves the best overall robustness, reducing the total number of performance regressions (Plateau + Rebound) by up to \robmax{\small$\times$}, 2.3{\small$\times$}, 1.8{\small$\times$} and 2.0{\small$\times$} compared to Bao, LOGER, Balsa, and LIMAO, respectively.
    \item \sysname exhibits enhanced efficiency through faster convergence. \sysname converges 1.1$\times$ faster on JOB (from 5.3 h to 4.7 h) and  2.4$\times$ on SSB (from 1.9 h to 0.8 h) compared to Balsa. Other baselines failed to converge, preventing direct comparison.
    \item In WRL, \sysname achieves 0.64 on JOB and 0.85 on SSB for the test set, indicating speedups of 1.55{\small$\times$} and 1.18{\small$\times$} compared to PostgreSQL. On TPC-DS, while other methods fail to reach PostgreSQL's performance, \sysname successfully catches up.
    \item We further validate the portability of \sysname on \commdb, where it maintains robust performance and achieves up to \effmax{\small$\times$} faster efficiency on the test set.
\end{itemize}

\subsection{Experimental Setup} \label{sec:setup}
\mypar{System setup} We conduct our experiments using PostgreSQL version 12.5. Specifically, PostgreSQL is set up with 32GB of shared buffers and cache size, 4GB of work memory. The Genetic Query Optimizer (GEQO) is disabled in all experiments to ensure compatibility with the \textit{pg\_hint\_plan} extension and to follow standard benchmarking practices. 
%\SL{Why disable GEQO?}
These settings closely align with prior work to ensure fair benchmarking against state-of-the-art query optimization techniques~\cite{JOB, Balsa}.

Our experiments utilize an NVIDIA RTX 6000 Ada GPU with 48GB of memory and Intel Xeon Gold 6530 CPUs for model training and inference.
% Our experiments utilize an NVIDIA RTX 6000 Ada GPU with 48GB of memory for model training and inference. The system is powered by Intel Xeon Gold 6530 CPUs, providing high computational throughput for both query execution and machine learning tasks.
The \commdb-based experiments are conducted under the same hardware configuration to ensure a fair comparison. We use the \commdb{} 2022 with default configuration settings. In this study, the \textit{expert plan} refers to the execution plan generated by the default cost-based optimizers of PostgreSQL and \commdb{}.
% , without any manual tuning.
%Microsoft SQL Server 2022 with default configuration settings, without any manual tuning.}
We experiment with Balsa as a base with our modules plugged in. All experiments are performed on a single agent in a nonparallel setting, and we use the median as the metric after 6 repetitions.

\begin{table}[h]
\centering
    \caption{Workload scale factor and train/test distribution.}
    \label{tab:workload}
    \begin{tabular}{c c c c c}
        \toprule
         & \textbf{Scale Factor} & \textbf{Queries} & \textbf{Train Set} & \textbf{Test Set} \\
        \midrule
        \textbf{JOB} & N/A & 113 & 87 & 26 \\
        \textbf{TPC-DS} & 4 & 57 & 42 & 15 \\
        \textbf{SSB} & 10 & 13 & 10 & 3\\
        \bottomrule
    \end{tabular}
%\vspace{-4mm}
\end{table}

\mypar{Datasets and workloads} We conduct experiments on three different datasets to evaluate the performance of \sysname: Join Order Benchmark~\cite{JOB}, TPC-DS~\cite{TPC-DS}, and Star Schema Benchmark~\cite{SSB}. Following the principle of rigorous and valid evaluation, we configure the workloads to have completely disjoint train and test sets at the template level. This strict isolation is essential to prevent performance inflation through structure leakage and to ensure the integrity of our evaluation metrics. By adhering to this standard of absolute separation, we verify that the performance gains of \sysname are derived from genuine structural robustness rather than simple pattern memorization. The split of train and test used in each benchmark is shown in \autoref{tab:workload}. %\JY{\sysname addresses the challenges of prior methods—specifically, slow convergence and weak robustness—while further enhancing overall performance. To comparatively evaluate its efficacy against prior approaches, we conducted experiments using the following workloads.}

\begin{figure*}[!t]
    \centering
    \subfloat[JOB]{
        \includegraphics[width=0.310\linewidth, height=4cm, keepaspectratio]{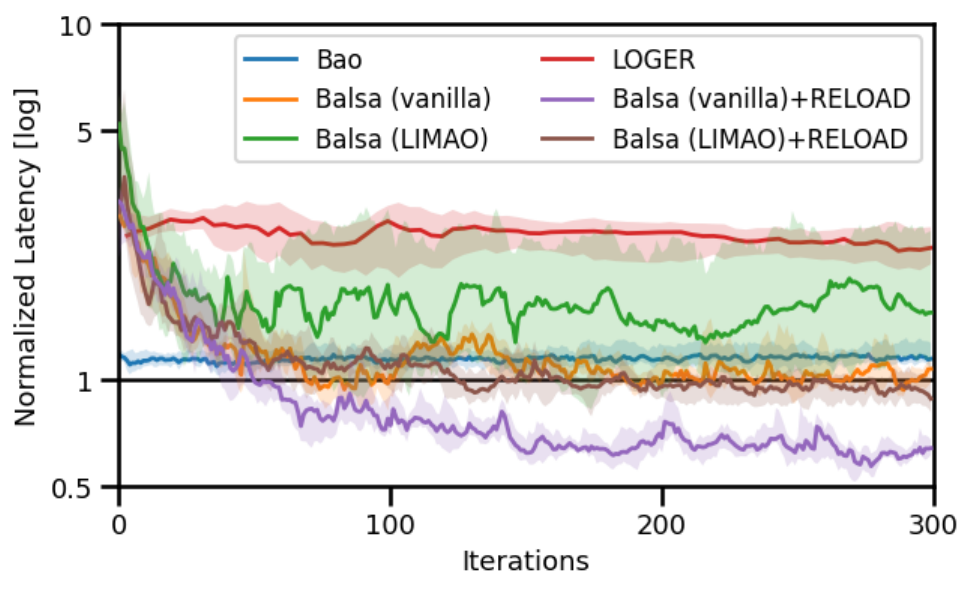}
        \label{macroTestJOB}
    }
    \hfill
    \subfloat[TPC-DS]{
        \includegraphics[width=0.310\linewidth, height=4cm, keepaspectratio]{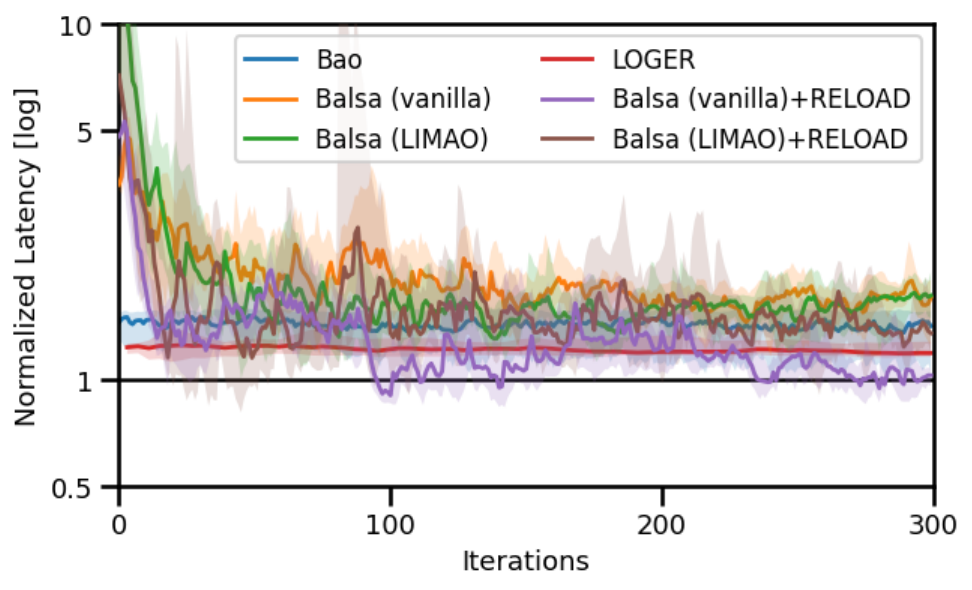}
        \label{macroTestTPCDS}
    }
    \hfill
    \subfloat[SSB]{
        \includegraphics[width=0.310\linewidth, height=4cm, keepaspectratio]{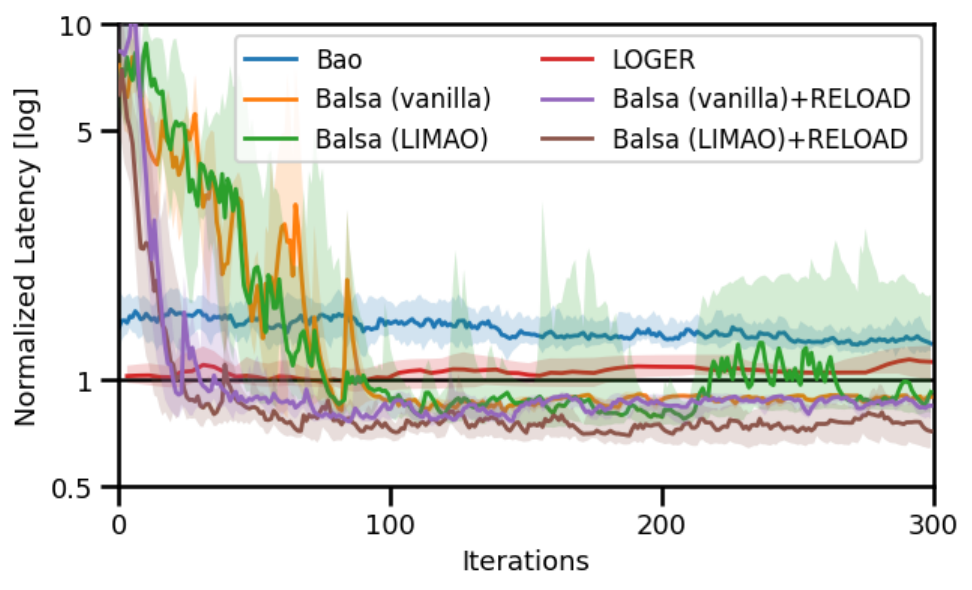}
        \label{macroTestSSB}
    }
    % \caption{Test performance of \sysname on PostgreSQL. Shaded areas denote variation across runs.}
    \caption{Performance of \sysname on test set in PostgreSQL. Shaded areas denote variation across runs.}
    \label{fig:macro_postgresql_test}
    % \vspace{-4pt}
\end{figure*}

\begin{figure*}[!t]
    \centering
    \subfloat[JOB]{
        \includegraphics[width=0.31\linewidth, height=4cm, keepaspectratio]{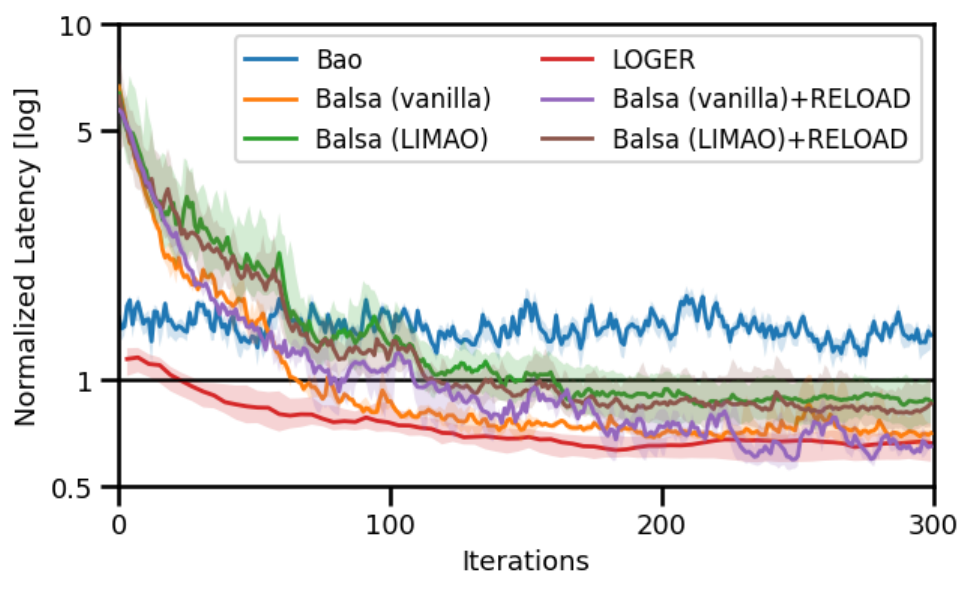}
        \label{macroTrainJOB}
    }
    \hfill
    \subfloat[TPC-DS]{
        \includegraphics[width=0.31\linewidth, height=4cm, keepaspectratio]{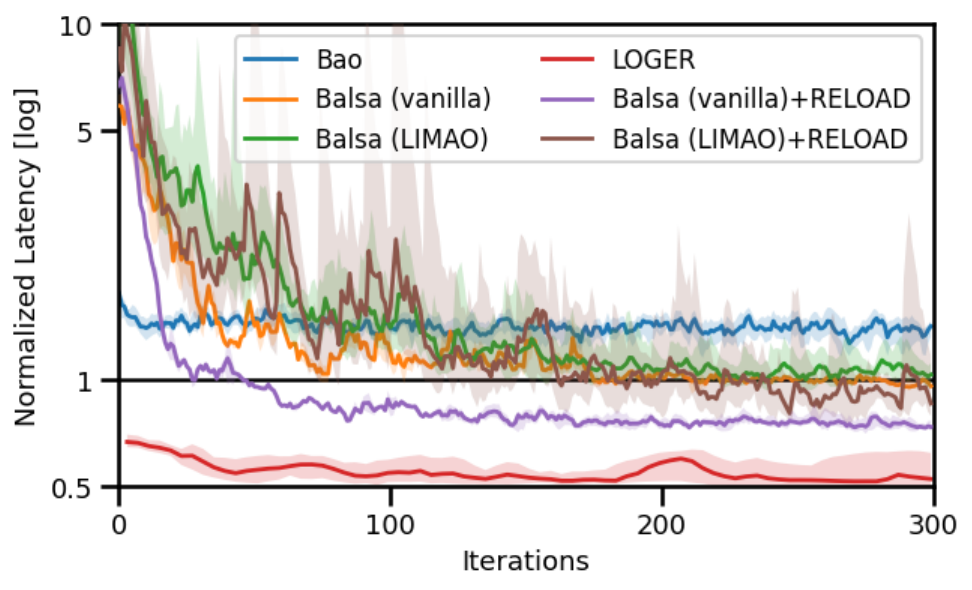}
        \label{macroTrainTPCDS}
    }
    \hfill
    \subfloat[SSB]{
        \includegraphics[width=0.31\linewidth, height=4cm, keepaspectratio]{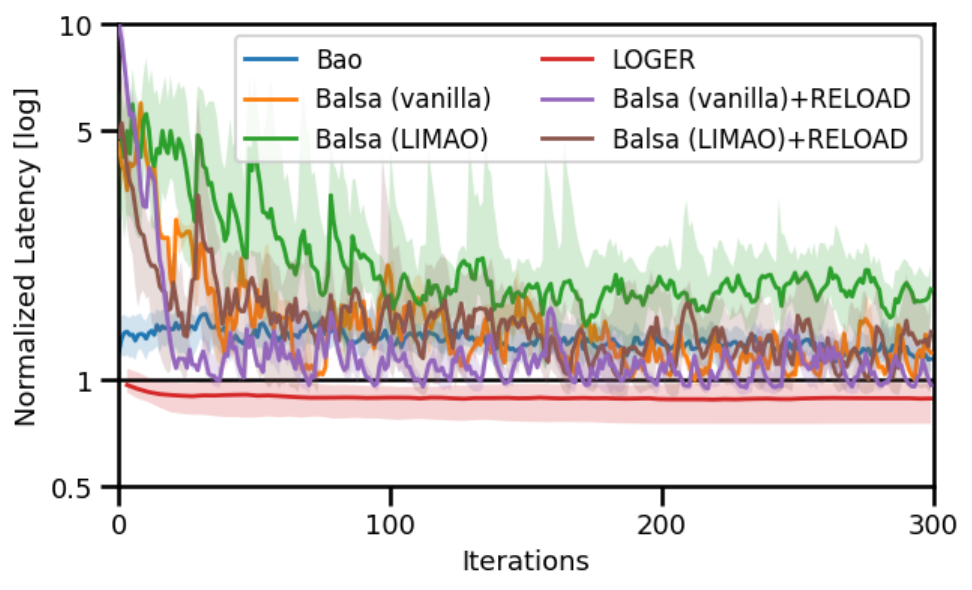}
        \label{macroTrainSSB}
    }
    % \caption{Training performance of \sysname on PostgreSQL. Shaded areas denote variation across runs.}
    \caption{Performance of \sysname on train set in PostgreSQL. Shaded areas denote variation across runs.}
    \label{fig:macro_postgresql_train}
    % \vspace{-4pt}
\end{figure*}

\textit{Join Order Benchmark (JOB)}: JOB is a real-world benchmark based on the IMDB dataset, containing 113 queries with complex joins and filters. We adopt the most challenging split, Base Query Sampling 3\footnote{The templates in the JOB, 1, 5, 12, 16, 22, 26, and 27, are used for testing.}, which assigns all variants of a query template to either train or test, preventing structure leakage~\cite{lqo-expect}.

\textit{TPC-DS}: TPC-DS is an industry standard benchmark, with tables based on the Snowflake schema~\cite{Why_TPCDS}. This benchmark contains 99 query templates. We have selected 19 select-project-join (SPJ) query templates and there are 3 generated queries for each query template. We ranked the top 5 templates based on Halstead complexity measures for each query and used them as our test set\footnote{The templates we use in TPC-DS are 3, 7, 12, 18, 20, 26, 27, 37, 42, 43, 52, 55, 62, 82, 84, 91, 96, 98, 99, and we used 12, 20, 43, 62, 99 for our test.}.

\textit{Star Schema Benchmark (SSB)}: SSB is simplified and optimized based on TPC-H and consists of 13 queries in total. We specifically selected SSB to evaluate join-heavy performance in a star schema environment. The queries are divided into four groups called “Query flight” and we use the most complex Q4 query flight, which joins all tables~\cite{variation-ssb}. This provides a more rigorous test of the optimizer's ability to navigate dense join paths than standard TPC-H.

\mypar{Baseline configuration}We compare \sysname with four baselines, each following their official setup. (1) Bao uses Thompson sampling over five arms—the configuration reported as optimal in its original paper. Each iteration includes 25 training queries, model update, and train/test evaluation. (2) LOGER uses a validation interval of 4; test latency is measured every 4 iterations. (3) Balsa runs in single-agent mode with the official simulator and minimal cost model for consistency. (4) LIMAO is implemented on top of Balsa and runs under the same experimental configurations to ensure a fair comparison.

\mypar{Implementation details} We set $k=256$, $\alpha=1$, $\beta=0.5$, $LR=10^{-3}$, and used 150 outer and 5 inner iterations for MAML. These values were determined through empirical tuning to ensure stable convergence.
% \mypar{Implementation details} \JY{For PER, we set the batch size to 256 with hyperparameters $\alpha=1$, $\beta=0.5$, and a learning rate of $LR=10^{-3}$. Regarding MAML, we used 150 outer and 5 inner iterations}. These values were determined through empirical tuning to ensure stable convergence.

% \yz{I really like the idea that RELOAD can be used jointly with existing RL algorihtm. I wonder if we want to emphasize this a bit more? Maybe in our archetecture, we can leave a room for thsoe "pluggable" RL algos. How do you think?}
\mypar{Robustness and efficiency criteria} We define robustness and efficiency by comparing each query's latency to that of an expert. For each query, the expert plan is executed ten times to estimate performance variability. The standard deviation of these executions is calculated, and twice the standard deviation (95\% confidence range under normal variability) is used as a tolerance band to account for measurement noise.
A query is considered superior or inferior based on whether its latency falls with this tolerance band. Two robustness failures are identified: Plateau (always inferior) and Rebound (initially superior, then regresses). Efficiency is the number of iterations needed for test latency to match the expert.

\subsection{End-to-end Evaluation} \label{sec:macro}
In this section, we compare the performance of \sysname against all state-of-the-art RL-based methods for each benchmark. We report overall performance using WRL, followed by robustness and efficiency analysis. Results on PostgreSQL, which demonstrate \sysname{}’s compatibility and improvements across all three metrics, are presented in \autoref{sec:end_postgresql}. Results on a commercial DBMS, which validate its portability under realistic system constraints, are presented in \autoref{sec:end_commdb}. In both cases, we use the same configurations: PER with a hybrid of recency and high TD error, and MAML with the best partitioning policy selected via DBI. 

\subsubsection{\sysname on PostgreSQL} \label{sec:end_postgresql} We first evaluate \sysname on %top of
the PostgreSQL query optimizer, one of the most %widely
used open-source DBMSs, to ensure compatibility with traditional database systems.

% \yz{maybe shrink the font size for this big table 4, or consider use bar chart}
\begin{table*}[!t]
\centering
\caption{Results of six methods---Bao, LOGER, Balsa (vanilla), Balsa (vanilla)+RELOAD, Balsa (LIMAO), and Balsa (LIMAO)+RELOAD---on three workloads. Robustness is shown as frequency/total, and efficiency as iterations (hours).
\label{tab:end_to_end_stat}
}
% three to five

% \yz{maybe bold the entry with the best perf for each column?}
\renewcommand{\arraystretch}{1.5} 
\setlength{\aboverulesep}{0ex}
\setlength{\belowrulesep}{0ex}
\setlength{\tabcolsep}{2.5pt} 

\begin{tabular}{c|
*{3}{>{\centering\arraybackslash}p{1.0cm}|>{\centering\arraybackslash}p{1.0cm}|>{\centering\arraybackslash}p{1.0cm}|}
>{\centering\arraybackslash}p{1.2cm}|
>{\centering\arraybackslash}p{1.2cm}|
>{\centering\arraybackslash}p{1.2cm}}
\toprule
\multirow{3}{*}{} & \multicolumn{9}{c|}{\textbf{Robustness}} & \multicolumn{3}{c}{\textbf{Efficiency}} \\ \cline{2-13} 
                  & \multicolumn{3}{c|}{JOB} & \multicolumn{3}{c|}{TPC-DS} & \multicolumn{3}{c|}{SSB} & JOB & TPC-DS & SSB  \\ \cline{2-13}
                  & Plateau & Rebound & Total & Plateau & Rebound & Total & Plateau & Rebound & Total &\multicolumn{3}{c}{Convergence time} \\ \hline
Bao               & 15/26 & 7/26 & 22/26 & 9/15 & 0/15 & 9/15 & 2/3 & 1/3 & 3/3 & NC & NC & NC \\ \hline
LOGER             & 18/26 & 3/26 & 21/26 & 4/15 & 1/15 & 5/15 & 1/3 & 1/3 & 2/3 & NC & NC & NC \\ \hline
Balsa (vanilla)   & 4/26 & 12/26 & 16/26 & 0/15 & 6/15 & 6/15 & 1/3 & 1/3 & 2/3 & 69(5.3) & NC & 77(1.9) \\ \hline
% \rowcolor{gray!20}
Balsa (vanilla)+RELOAD & 5/26 & 4/26 & \textbf{9/26} & 0/15 & 4/15 & \textbf{4/15} & 0/3 & 2/3 & 2/3 & \textbf{50(4.7)} & \textbf{95(1.3)} & \textbf{20(0.8)} \\ \hline
Balsa (LIMAO)     & 8/26 & 9/26 & 17/26 & 0/15 & 8/15 & 8/15 & 0/3 & 2/3 & 2/3 & NC & NC & 76(4.9) \\ \hline
% \rowcolor{gray!20}
Balsa (LIMAO)+RELOAD & 7/26 & 5/26 & 12/26 & 0/15 & 8/15 & 8/15 & 0/3 & 1/3 & \textbf{1/3} & 125(16.7) & NC & 21(0.8) \\ 
\specialrule{1.2pt}{0pt}{0pt}
\raisebox{-0.6ex}{\textbf{Best}} & \multicolumn{3}{c|}{\raisebox{-0.8ex}{\makecell[c]{Balsa (vanilla)\\+RELOAD}}}
& \multicolumn{3}{c|}{\raisebox{-0.8ex}{\makecell[c]{Balsa (vanilla)\\+RELOAD}}}  
& \multicolumn{3}{c|}{\raisebox{-0.8ex}{\makecell[c]{Balsa(LIMAO)\\+RELOAD}}}  
& \multicolumn{3}{c}{\raisebox{-0.8ex}{\makecell[c]{Balsa (vanilla)\\+RELOAD}}} \\
\bottomrule
\end{tabular}

\raggedright
\vspace{1mm}
% \hspace{1cm}
\textit{Note:} NC = No Convergence. 
% \JY{$^{*}$ = 2 out of 6 runs did not converge (NC).}
\par
% \vspace{-3mm}
\end{table*}

\mypar{Performance Overview} 
% \yz{can you explain each line? there are solid and dashed ones. Maybe separate into two figures. it was a bit too many lines in one. 10?}
\autoref{fig:macro_postgresql_test} and \autoref{fig:macro_postgresql_train} show the performance of our experiments in three workloads. Although WRL is not our primary metric, it has been used as a performance indicator in previous studies. After 300 iterations, \sysname achieves WRL of 0.65, 0.73, and 0.88 on the train set for JOB, TPC-DS, and SSB, corresponding to speedups of 1.54{\small$\times$}, 1.37{\small$\times$}, and 1.13{\small$\times$}, respectively. On the test set, RELOAD achieves 0.64, 1.02, and 0.85 in WRL, demonstrating speedups of 1.55{\small$\times$}, 0.98{\small$\times$}, and 1.18{\small$\times$}, respectively. 
Results are reported for the Balsa (vanilla)+\sysname configuration, representative of overall \sysname performance.
Across all workloads, Bao consistently shows the weakest robustness, as it fails to outperform PostgreSQL and struggling to maintain consistent performance across various queries.
LOGER achieves rapid convergence on the training set but shows limited stability on queries not encountered during training.
Balsa performs competitively on JOB and SSB but underperforms on TPC-DS. 
LIMAO performs moderately but remains sensitive to query templates excluded from the training set.
When combined with \sysname, both Balsa and LIMAO exhibit faster convergence and stronger robustness, confirming \sysname{}’s effectiveness.

\begin{figure*}[!t]
\vspace{-8mm}
  \centering
  \begin{minipage}[t]{0.67\textwidth}
  \vspace{0pt}
    \centering
    \subfloat[{\fontsize{6.5}{6.5}\selectfont Bao}]{
        \includegraphics[width=0.153\linewidth]{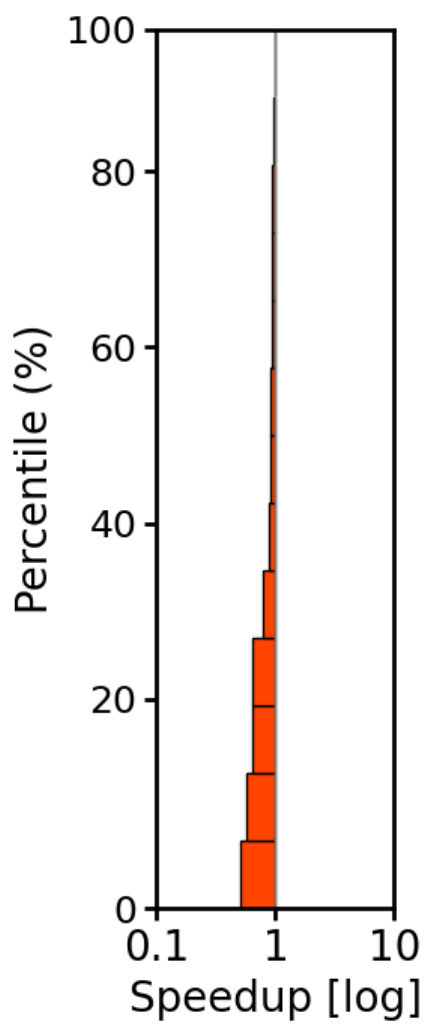}
        \label{speedupBao}
    }
    \subfloat[{\fontsize{6.5}{6.5}\selectfont LOGER}]{
      \includegraphics[width=0.153\linewidth]{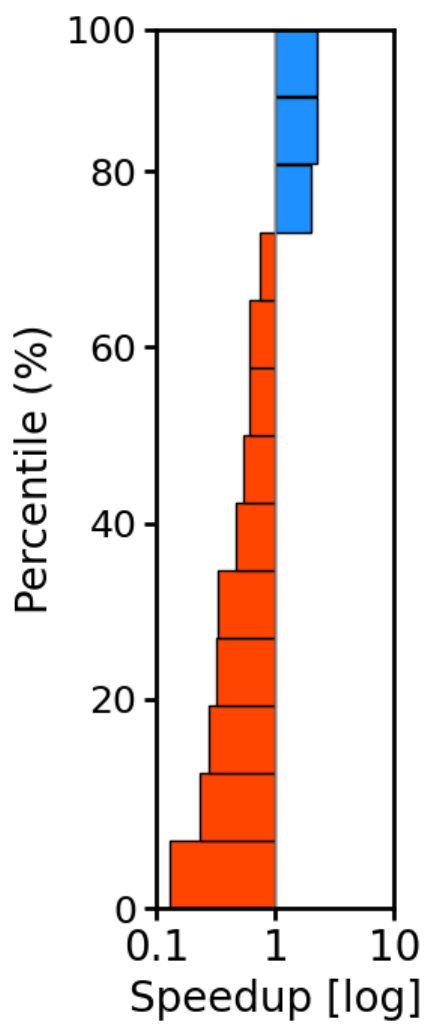}
      \label{speedupLOGER}
    }
    \subfloat[{\fontsize{6.5}{6.5}\selectfont Balsa (vanilla)}]{
      \includegraphics[width=0.153\linewidth]{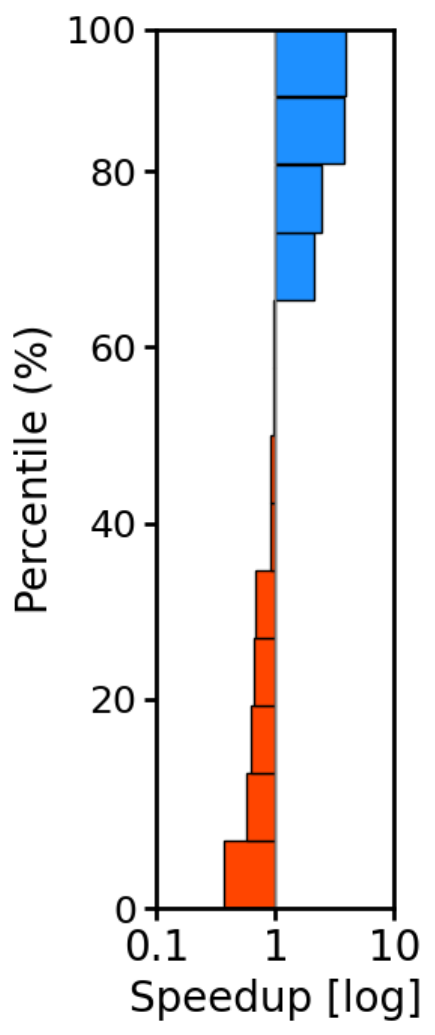}
      \label{speedupBalsaV}
    }
    \subfloat[{\fontsize{6.5}{7.5}\selectfont {\parbox[t]{1.6cm}{\centering Balsa (vanilla) \\ +RELOAD}}}]{
      \includegraphics[width=0.153\linewidth]{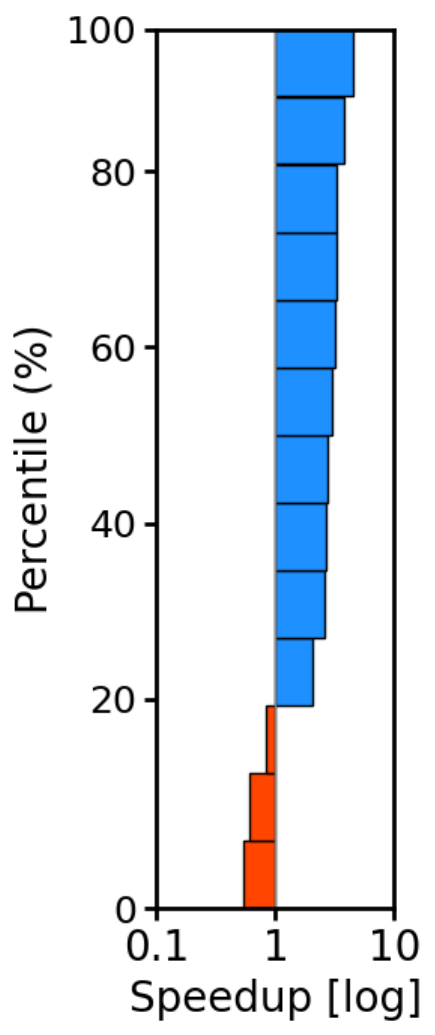}
      \label{speedupBalsaReload}
    }
    \subfloat[{\fontsize{6.5}{6.5}\selectfont Balsa (LIMAO)}]{
      \includegraphics[width=0.153\linewidth]{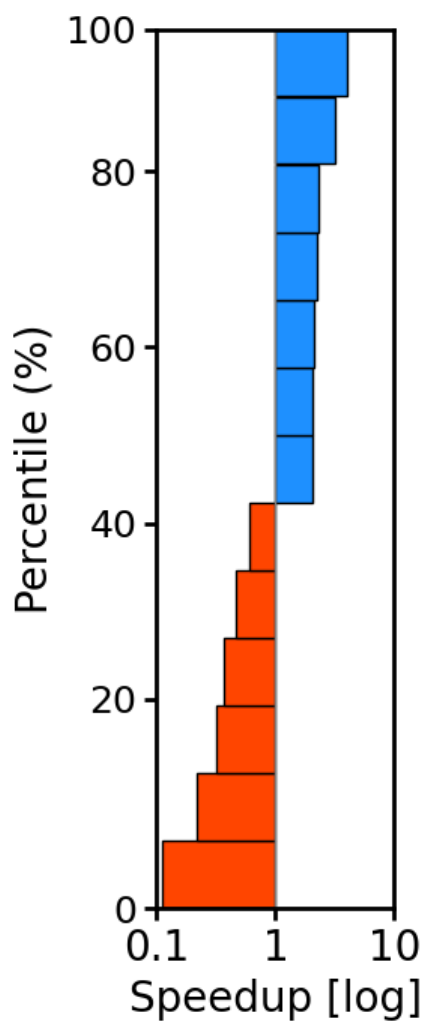}
      \label{speedupLimao}
    }
    \subfloat[\fontsize{6.5}{7.5}\selectfont {\parbox[t]{1.6cm}{\centering Balsa (LIMAO) \\ +RELOAD}}]{
        \includegraphics[width=0.153\linewidth]{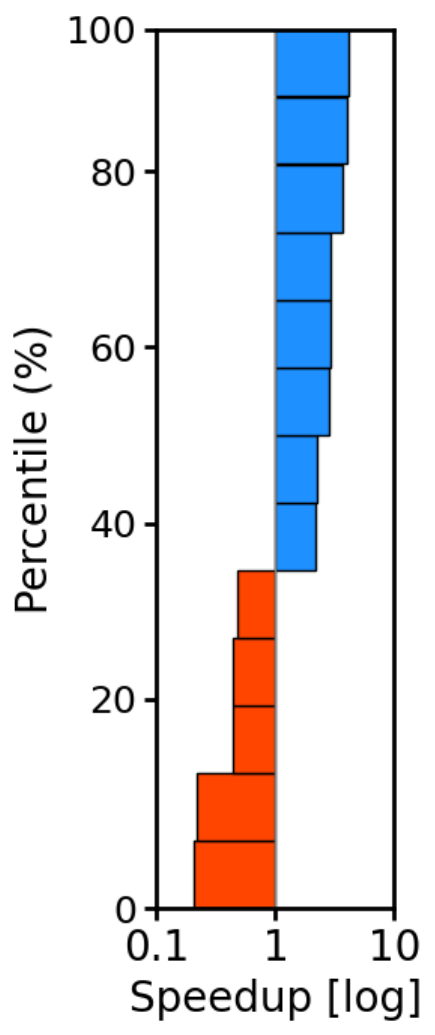}
        \label{speedupLimaoReload}
    }
    \caption{Speedup in execution latency per test query ($>$500ms
        with PostgreSQL's plan) in JOB. A total of 13 test queries satisfying this condition are shown.}
    \label{fig:robustness_per_query}
  \end{minipage}
  \hfill
  \begin{minipage}[t]{0.30\textwidth}
  \vspace{0pt}
    \centering
    \vspace{10pt}
    \includegraphics[width=\linewidth]{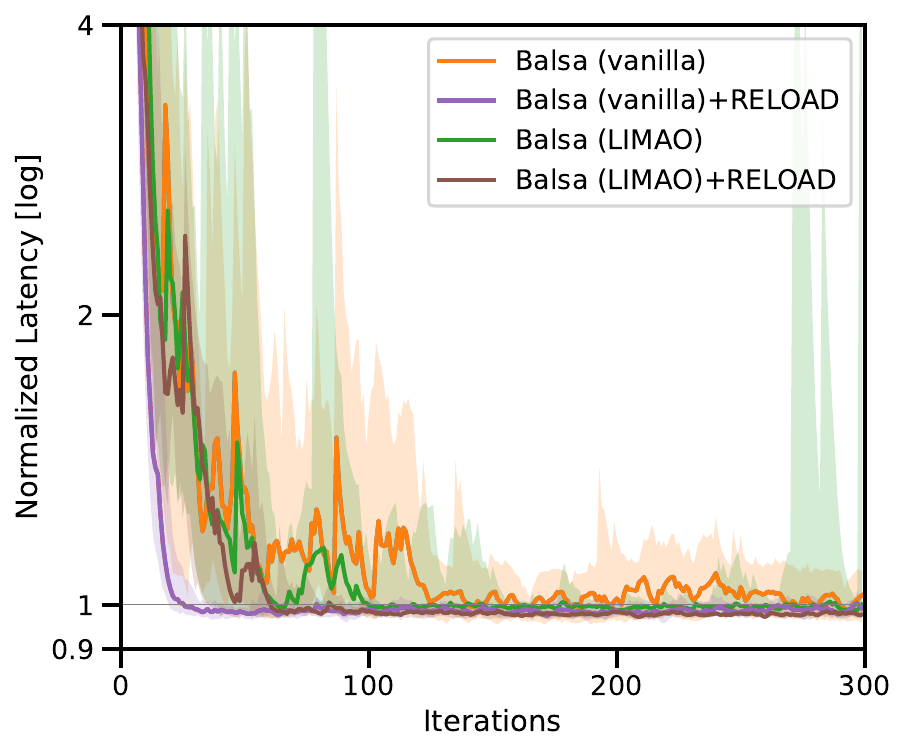}
    % \vspace{1pt plus 0fill}
    \caption{Impact of \sysname on \commdb under the SSB test set. Shaded areas denoted variation across runs.}
    \label{fig:macro_sqlserver}
  \end{minipage}
\vspace{-6mm}
\end{figure*}

% \begin{figure*}[!t]
%     \centering
%     \begin{subfigure}{0.47\linewidth}
%         \centering
%         \includegraphics[width=\linewidth, height=5cm, keepaspectratio]{figures/evaluation/micro/per_overall.pdf}
%         \caption{Impact of weighting policies on \buffer.}
%         \label{fig:micro:per}
%     \end{subfigure}
%     \hfill
%     \begin{subfigure}{0.47\linewidth}
%         \centering
%         \includegraphics[width=\linewidth, height=5cm, keepaspectratio]{figures/evaluation/micro/maml_overall.pdf}

% \caption{Impact of partitioning polices on \maml.}
% \label{fig:micro:maml}
%     \end{subfigure}
%     \hfill
%     \caption{Micro-experiment results of \sysname on the JOB test set.
% (a) evaluates the impact of different weighting strategies in knowledge retention using PER, and (b) compares query and data complexity–based partitioning policies for knowledge transfer with MAML. Vertical dashed lines indicate the convergence point of each strategy. Results are reported as the median of six runs. Both experiments examine module variants, including baselines without retention or transfer, to evaluate their impact on robustness and efficiency.}
%     \label{fig:micro}
%     % \vspace{-4mm}
% \end{figure*}

\mypar{Robustness} 
The results are summarized in \autoref{tab:end_to_end_stat}. Robustness is evaluated by counting the total number of Plateaus and Rebounds observed during training, where smaller values indicate higher stability and fewer performance regressions.
On JOB, \sysname substantially improves robustness. Balsa (vanilla)+\sysname reduces the number of Plateaus and Rebounds from 16 to 9, a 44\% reduction compared with the baseline. Similarly, when integrated with LIMAO, \sysname reduces the total number of performance regressions by about 30\%, demonstrating that it stabilizes training even when applied to an optimizer originally designed for online adaptation. As shown in \autoref{fig:robustness_per_query}, \sysname further improves query-level performance, yielding speedups for about 77\% of test queries with Balsa (vanilla)+\sysname, and roughly 62\% with Balsa (LIMAO)+\sysname, despite LIMAO’s stronger baseline.
On TPC-DS , \sysname with Balsa (vanilla) achieves the lowest total regression count 4. When paired with LIMAO, \sysname maintains a similar level of robustness. This is likely because LIMAO's predefined k-prototypes do not fully align with the characteristics of the TPC-DS workload, thereby limiting additional improvement.
For SSB, the smaller query set makes performance differences less pronounced, resulting in generally comparable performance across all methods.

Specifically, on JOB, all methods commonly struggle with template q1. LOGER shows weak robustness in templates q12, q17, and q22, while Balsa shows weak robustness in templates q5, q22, and q28.
On TPC-DS, LOGER struggles with templates query43, query62, and query99; Balsa struggles with templates query12, query62, and query99; and \sysname struggles with template query20. In particular, query20 exhibits an opposite performance trend compared to query62 and query99, implying that they have significantly different strategies for generating optimal query plans.

We next analyze how each method behaves across query templates to explain these robustness differences.
Bao, unlike these methods that fail only on specific templates, struggles across all query templates and workloads because its online update scheme lacks a fixed training set, limiting its ability to maintain consistent performance per query.
LOGER achieves fast and stable convergence by restricting its search space for plan operators, which improves training stability but shows limited reliability on query patterns not observed during training.
Balsa achieves solid performance, but occasionally shows plateau or rebound behaviors on certain templates.
LIMAO, designed for online adaptation based on recurring subpatterns, shows limited robustness under the template-disjoint setting.
In contrast, \sysname consistently improves robustness across both Balsa (vanilla) and Balsa (LIMAO), mitigating template-specific regressions while maintaining stable convergence behavior.
These results demonstrate that \sysname substantially improves performance stability across a broad range of individual queries. By prioritizing informative, underlearned, and timely experiences, the knowledge retention module effectively mitigates plateau and rebound behaviors. 
It further mitigates the underlying local optima and credit assignment issues that have long hindered stability in RL-based query optimization.

\mypar{Efficiency} 
As shown in \autoref{tab:end_to_end_stat}, \sysname consistently accelerates convergence across all workloads. On JOB, Balsa (vanilla)+RELOAD converges in 50 iterations (4.7 h), faster than the baseline Balsa. On TPC-DS, \sysname is the only configuration that successfully converges, completing in 95 iterations (1.3 h).
On SSB, both \sysname variants reach convergence within an hour, while all baselines require longer or fail to converge. 
These results indicate that existing RL-based optimizers often suffer from slow convergence due to the absence of cross-task knowledge transfer, requiring extensive interactions. 
In contrast, \sysname mitigates this inefficiency by initializing value model with meta-learned parameters. 
The integration of knowledge transfer substantially reduces the learning overhead, allowing \sysname to shorten the initial ramp-up phase and achieve rapid, consistent convergence across diverse query patterns.
% This enables the optimizer to adapt faster to new queries, achieving faster and more consistent convergence across diverse workloads.

\subsubsection{\sysname on \commdb} \label{sec:end_commdb} To verify the competitiveness of \sysname on commercial database systems, we conducted experiments on \commdb. As the JOB and TPC-DS workloads exhibited trends consistent with the PostgreSQL results, we report only SSB for clarity and space efficiency. 
We adapted the query hints to conform to \commdb syntax, ensuring that the queries would execute correctly and preserve their intended optimization behavior.

\mypar{Performance Overview} \autoref{fig:macro_sqlserver} shows the experimental results of \sysname on the SSB workload running on \commdb. While \commdb incorporates advanced system-level optimizations, \sysname still delivers consistent improvements in WRL---1.12$\times$ on the training set and 1.01$\times$ on the test set.

\mypar{Robustness} All configurations show stable performance, except for a single Rebound in Balsa (LIMAO). Both \sysname variants maintain robust performance across all test queries, even within a commercial database system.

\mypar{Efficiency} \sysname achieves efficiency by 3.1$\times$ on Balsa (vanilla) (from 0.4 h to 0.13 h) and 1.4$\times$ on Balsa (LIMAO) (from 2.1 h to 1.5 h). These results confirm that \sysname consistently improves efficiency across both optimizers.

\subsection{Micro-experiments} \label{sec:micro}
In this section, we analyze how different design choices affect robustness and efficiency. 
For the PER module (\autoref{sec:micro_per}), we test five configurations: four proposed weighting policies—recency, low TD error, high TD error, and their combination—and one ablation without knowledge retention.
For the MAML module (\autoref{sec:micro_maml}), we also evaluate five configurations: four partitioning policies—Halstead complexity, total number of operators, estimated query cost, and estimated rows—and one ablation without knowledge transfer. 
All experiments are performed on the JOB workload.

\begin{figure*}[!t]
    \centering
    \subfloat[Impact of weighting policies on \buffer.]{
        \centering
        \includegraphics[width=0.47\linewidth, height=5cm, keepaspectratio]{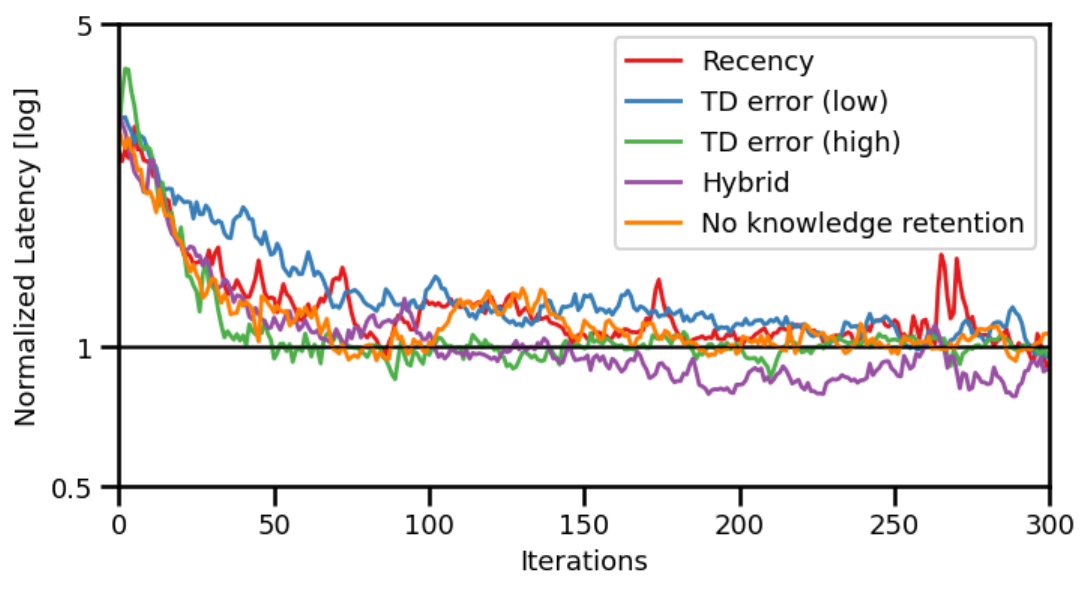}
        \label{fig:micro:per}
    }
    \hfill
    \subfloat[Impact of partitioning polices on \maml.]{
        \centering
        \includegraphics[width=0.47\linewidth, height=5cm, keepaspectratio]{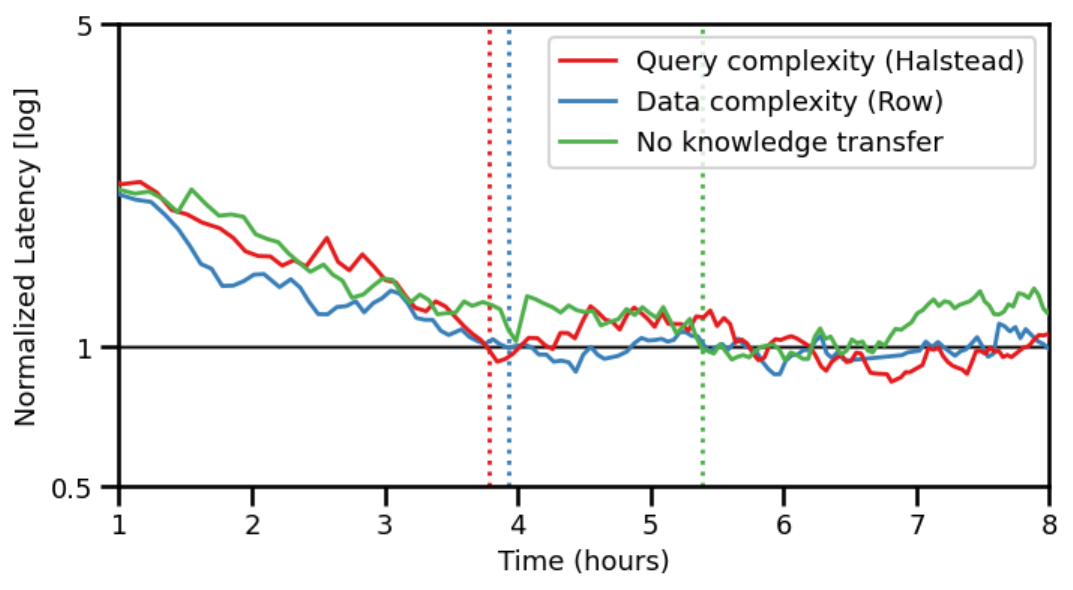}
    \label{fig:micro:maml}
    }
    \caption{Micro-experiment results of \sysname on the JOB test set.
(a) evaluates the impact of different weighting strategies in knowledge retention using PER, and (b) compares query and data complexity–based partitioning policies for knowledge transfer with MAML. Vertical dashed lines indicate the convergence point of each strategy. Results are reported as the median of six runs. Both experiments examine module variants, including baselines without retention or transfer, to evaluate their impact on robustness and efficiency.}
    \label{fig:micro}
    % \vspace{-4mm}
\end{figure*}

\subsubsection{\buffer} \label{sec:micro_per}
We first analyze the impact of \buffer on robustness. \autoref{fig:micro:per} compares five configurations, performing PER with different weight policies for each.

The baseline without knowledge retention shows the largest fluctuations, confirming that learning solely from recent samples leads to unstable behavior. 
Recency-based sampling also exhibits high variability throughout training and fails to consistently approach the expert performance, indicating that emphasizing only recent experiences results in short-term bias without meaningful improvement.
TD error (high) reaches the expert performance early but struggles to improve further, suggesting that emphasizing only large TD errors helps accelerate initial convergence yet limits long-term progress.
In contrast, TD error (low) improves steadily and maintains stable performance, but its slower learning pace causes it to catch up to the expert only near the end.
Among all policies, the Hybrid policy---combining recency and high TD error---achieves the best balance, showing stable progress and maintaining latency consistently below the expert baseline.
This combination implicitly mitigates credit assignment issues by reinforcing meaningful states and helps the agent escape local optima caused by overfitting to recent or easy samples.
Recency ensures continuous policy refinement, while high TD error emphasizes underlearned but valuable experiences, leading to more balanced and reliable performance across all query patterns.
As a result, \sysname achieves stable robustness, confirming the effectiveness of experience prioritization through PER weighting.

\subsubsection{\maml} \label{sec:micro_maml} We investigate the effect of \maml, which is key to improving efficiency. We evaluate five configurations and perform MAML for each grouping.
\autoref{fig:micro:maml} presents representative results on the JOB test set.
For clarity, we show only the Halstead-based partitioning policy among the query complexity configurations, as others exhibit similar trends. This policy also achieved the smallest DBI, consistent with its fastest and most stable convergence observed in the figure.
All MAML-based configurations converge faster and more stably than the baseline without knowledge transfer, confirming that cross-task initialization enhances training efficiency. 
However, none of the configurations outperforms expert-level performance, indicating that faster convergence alone does not ensure optimal learning outcomes. 
Overall, these results highlight that while meta-learning accelerates learning, effective task grouping remains essential for maintaining stable training.
% and strong generalization to unseen queries. \JY{last sentence...}
\section{Related work}

%%%%% Version 1 %%%%%
\mypar{Learned query optimizers} 
Learned query optimizers use ML to overcome traditional optimizers’ reliance on static heuristics and cost models, improving adaptability to complex query workloads \cite{LQOSurvey}.
A common theme across prior work is to enhance plan quality by improving plan representations, exploration strategies, or cost/value estimation models. For example, Neo~\cite{NEO}, ReJoin~\cite{ReJOIN} and Balsa~\cite{Balsa} employ deep reinforcement learning to generate execution plans, while Bao~\cite{Bao_QO} guides plan selection among candidates generated by the underlying DBMS using learned value models. 
GLO~\cite{chen2024glo} integrates DBMS statistics and Transformer-based value models, and Athena~\cite{li2025athena} diversifies join-order candidates and applies a learned comparator for plan selection.
While these approaches consistently demonstrate performance improvements at the workload level, their training objectives—typically formulated to minimize aggregate loss or maximize expected reward over a workload—implicitly.
In contrast, \sysname{} explicitly targets query-level robustness and learning efficiency, addressing performance regressions that hinder practical adoption of learned optimizers in commercial database systems.

\mypar{Addressing performance regression and suboptimality}
Performance regression can occur in certain machine learning scenarios, particularly when training iteratively in nonstationary environments. In query optimizers, mitigating regression and addressing suboptimality are crucial for achieving robust optimization.
Lero~\cite{Lero} enhances stability by learning relative rankings between subqueries instead of predicting absolute latencies, improving robustness. 
LOGER~\cite{LOGER} and Eraser~\cite{Eraser} achieve robustness by deliberately narrowing the plan selection space, encouraging the optimizer to favor more stable execution plans.
In contrast to these methods, \sysname introduces specialized PER, which prioritizes experiences based on recency and TD error, effectively mitigating regression, preventing premature convergence, and enhancing adaptability in complex optimization tasks.

\mypar{Knowledge retention}
In query optimization, preserving past knowledge is essential for maintaining stable performance under evolving workloads. LIMAO~\cite{LIMAO} addresses this by introducing a lifelong learning framework that stores policies in a module hub and reuses them when workloads drift, demonstrating the feasibility of continual learning. However, its workload-level design leads to coarse-grained adaptation and limited rapid convergence. %robustness when facing unseen templates. 
\sysname extends this direction by adopting a sampling-based approach that prioritizes fine-grained sub-plan experiences through PER rather than reusing clustered modules as in LIMAO. In addition to knowledge retention, \sysname incorporates a knowledge transfer mechanism based on meta-learning, enabling efficient convergence to surpass expert-level performance.

\section{Conclusion}
In this paper, we propose \sysname, a robust and efficient learned query optimizer for advanced database systems.
\sysname enhances robustness and efficiency in RL-based optimizers through two complementary modules:
(1) \textit{knowledge retention}, which employs PER to overcome local optima and alleviate credit assignment issues under sparse rewards, and
(2) \textit{knowledge transfer}, which applies MAML for rapid adaptation via complexity-aware task grouping.
Experiments on JOB, TPC-DS, and SSB show that \sysname significantly mitigates performance regressions and accelerates convergence to expert-level performance, improving robustness by up to \robmax{\small$\times$} and efficiency by up to \effmax{\small$\times$}. Overall, \sysname provides a practical and general framework for robust and efficient learned query optimization, serving as a foundation for more adaptive and reliable optimizers.

%In this paper, we propose \sysname, a robust and efficient learned optimizer adapting to dynamic workloads. Robustness, defined as the ability to minimize performance regression such as Plateau and Rebound, and efficiency, measured by convergence time, are critical challenges. \sysname addresses the limitations of existing RL-based query optimization methods to ensure consistent performance in diverse workloads through two main components: knowledge retention and knowledge transfer. Knowledge retention uses PER to prevent catastrophic forgetting and improve robustness by focusing on key training experiences. Knowledge transfer applies MAML for faster adaptation and convergence, balancing generalization and customization through complexity-aware task partitioning.

%Experiments on JOB, TPC-DS, and SSB demonstrated \sysname’s advantage over methods like Balsa and LOGER, consistently achieving better robustness and faster convergence, even in cases where others failed. For instance, \sysname improves robustness by up to 3.08x and successfully converges on all workloads. In summary, RELOAD combines robustness, efficiency, and adaptability, marking significant progress in query optimization and demonstrating potential for real-world applications. Future work could extend RELOAD towards lifelong RL to continuously adapt and optimize for more complex scenarios and evolving workloads.
\section*{AI-Generated Content Acknowledgement}
We utilized ChatGPT throughout the manuscript to assist with minor language proofreading and grammar refinement. Additionally, ChatGPT provided limited support in improving the visualization quality of figures in Chapter 6 (Evaluation) through Python code refinement. All AI-assisted content was carefully reviewed, verified, and edited by the authors. The technical ideas, analysis, and results presented in this paper are entirely the work of the authors.

\bibliographystyle{IEEEtran}
\bibliography{reference}
\end{document}